\documentclass[graybox]{svmult}

\usepackage{mathptmx}       
\usepackage{helvet}         
\usepackage{courier}        
\usepackage{type1cm}        
\usepackage{makeidx}         
\usepackage{graphicx}        
\usepackage{multicol}        
\usepackage[bottom]{footmisc}

\usepackage{amsmath,amssymb}

\makeindex
\usepackage{bm}
\usepackage{float}
\restylefloat{figure}
\usepackage{shuffle}
\usepackage{array}
\usepackage{bbm}

\restylefloat{figure}

\usepackage{epstopdf}

\newcommand{\req}[1]{(\ref{#1})}
\def\vev#1{\langle #1 \rangle}

\def\fc#1#2{\frac{#1}{#2}}
\def\h{\frac{1}{2}}
\newcommand{\nwc}{\newcommand}
\nwc{\ba}  {\begin{array}}
\nwc{\ea}  {\end{array}}
\nwc{\bdm} {\begin{displaymath}}
\nwc{\edm} {\end{displaymath}}

\nwc{\bea} {\begin{equation}\ba{lcl}}
\nwc{\eea} {\ea\end{equation}}

\nwc{\be} {\begin{equation}}
\nwc{\ee} {\end{equation}}

\nwc{\bda} {\bdm\ba{lcl}}
\nwc{\eda} {\ea\edm}

\nwc{\bc}  {\begin{center}}
\nwc{\ec}  {\end{center}}

\nwc{\ds}  {\displaystyle}

\nwc{\nn} {\nonumber}
\nwc{\nnn} {\nonumber \vspace{.2cm} \\ }
\nwc{\ra}{\rightarrow}
\nwc{\lra}{\longrightarrow}
\def\lf{\left}\def\ri{\right}

\nwc{\p} {\partial}
\nwc{\Tr}{{\rm Tr}}
\def\IR{{\bf R}}
\def\Fc{{\cal F}}
\def\ap{\alpha'}

\def\Mc{{\cal M}}

\def\ov{\overline}

\def\eps{\epsilon}
\def\z{\zeta}
\def\al{\alpha}

\def\Oc{{\cal O}}

\def\IQ{{\bf Q}}
\def\IN{{\bf N}}
\def\IR{{\bf R}}
\def\IC{{\bf C}}
\def\IP{{\bf P}}
\def\IZ{{\bf Z}}

\def\Qc{{\mathcal{Q}}}
\def\Af{{ \frak{A} }}
\def\Ac{{\cal A}}
\def\Lc{{\cal L}}
\def\Fc{{\cal F}}
\def\Gc{{\cal G}}

\def\eps{\epsilon}
\def\al{\alpha}

\def\si{\sigma}\def\Si{{\Sigma}}

\def\Mo{{\cal M}_{0,N}}
\def\sv{{\rm sv}}
\def\SVM{{\zeta^{\frak m}_{\rm sv}}}
\def\SV{{\zeta_{\rm sv}}}
\def\ad{{\rm ad}}
\def\Uc{{\cal U}}
\def\Zc{{\cal Z}}
\def\Hc{{\cal H}}
\def\per{\rm per}

\def\MZ#1{\zeta^{\frak m}_{#1}}

\makeindex             

\def\tfrac{\fc}
\def\bm{\vec}

\begin{document}

\title*{Periods and Superstring Amplitudes}
\author{S. Stieberger}
\institute{S. Stieberger \at Max--Planck--Institut f\"ur Physik,Werner--Heisenberg--Institut, 80805 M\"unchen, Germany \\ \email{stephan.stieberger@mpp.mpg.de}}
%
%
\maketitle

\abstract{Scattering amplitudes which describe the interaction of physical states play an important role in determining physical observables. In string theory the physical states are given by vibrations of open and closed strings and their interactions are described (at the leading order in perturbation theory) by a world--sheet given by the topology of a disk or sphere, respectively. Formally, for scattering of $N$ strings this leads to $N\!-\!3$--dimensional iterated real integrals along the compactified real axis or $N\!-\!3$--dimensional  complex sphere integrals, respectively.
As a consequence the physical observables are described by periods on $\Mo$ -- the moduli space of Riemann spheres of $N$ ordered marked points.
The mathematical structure of these string amplitudes share many recent advances in arithmetic algebraic geometry and number theory like multiple zeta values, single--valued multiple zeta values, Drinfeld, Deligne associators, Hopf algebra and Lie algebra structures related to Grothendiecks Galois theory.
We review these results, with emphasis on a beautiful link between generalized hypergeometric functions describing the real iterated integrals on $\Mo(\IR)$ and the decomposition of motivic multiple zeta values. Furthermore, a relation expressing complex integrals on $\Mo(\IC)$  as single--valued projection of iterated real integrals  on 
$\Mo(\IR)$  is exhibited.}

\section{Introduction}
\label{sec:1}

During the last years a great deal of work has been addressed to the problem of revealing and understanding the hidden mathematical structures of scattering amplitudes in both field-- and string theory.
Particular emphasis on their underlying geometric structures seems to be especially 
fruitful and might eventually yield an alternative\footnote{In field--theory with ${\cal N}=4$ supersymmetry such methods have recently been pioneered by using tools in algebraic geometry \cite{ArkaniHamedNW,ArkaniHamedJHA} and arithmetic algebraic geometry
\cite{Goncharov,GoldenXVA}.} way of constructing perturbative amplitudes by methods residing in  arithmetic algebraic geometry
.
In such a framework physical quantities are given by periods (or more generally by $L$--functions) typically describing the volume of some polytope or integrals of a discriminantal configuration (a configuration of multivariate hyperplanes).
The mathematical quantities which occur in string amplitude computations are periods
which relate to fundamental objects in number theory and  algebraic geometry. A period is a complex number whose real and imaginary parts are given by absolutely convergent integrals of rational functions with rational coefficients
over domains in $\IR^n$ described by polynomial inequalities with rational coefficients.
More generally, periods are values of integrals of algebraic  differential forms over certain chains in algebraic varieties~\cite{Kontsevich}.
E.g.~in quantum field theory  the coefficients of the Laurent series in the parameter 
$\eps=\h(4\!-\!D)$  of dimensionally regulated Feynman integrals are numerical periods in the Euclidian region with all ratios of invariants and masses having rational values \cite{BognerMN}. Furthermore, the power series expansion in the inverse string tension $\ap$ of tree--level superstring amplitudes yields iterated integrals \cite{OprisaWU,StiebergerTE,BroedelTTA}, 
which are periods of the moduli space $\Mc_{0,N}$ of genus zero curves with $N$ ordered marked points 
\cite{GM} and integrate to $\IQ$--linear combinations of multiple zeta values (MZVs) 
\cite{BrownQJA,Terasoma}. Similar considerations \cite{BL} are
expected to hold at higher genus in string perturbation theory, cf.~\cite{BroedelVLA} for some recent investigations at one--loop.
At any rate, the analytic dependence on the inverse string tension~$\ap$
of string  amplitudes furnishes an extensive and rich mathematical structure, which is
suited to exhibit and study modern developments in number theory and arithmetic algebraic geometry.

The forms and chains entering the definition of periods may depend on parameters (moduli). As a consequence the periods satisfy linear differential equations with algebraic coefficients.  This type of
differential equations is known as generalized Picard--Fuchs equations or Gauss--Manin systems. A subclass of the latter describes the $A$--hypergeometric 
system\footnote{The initial data for a GKZ--system is an integer matrix $A\in \IZ^{r\times n}$ 
together with a parameter vector~$\gamma\in\IC^r$. For a given matrix $A$ the structure of the GKZ--system depends on the properties of the vector~$\gamma$ defining non--resonant and resonant systems. E.g.~a non--resonant system of $A$--hypergeometric equations is irreducible~\cite{Beuk}.} or Gelfand--Kapranov--Zelevinsky (GKZ) system relevant to tree--level string scattering. 
One notorious example of periods are multivariate (multidimensional) or generalized hypergeometric functions\footnote{More precisely, at an algebraic value of their argument their value  is $\fc{1}{\pi}\wp$, with $\wp$ being the set of periods.}. 
In the non--resonant case the solutions of the GKZ system can be represented by 
generalized Euler integrals \cite{GKZ}, which appear as world--sheet integrals in superstring tree--level amplitudes and integrate to multiple Gaussian hypergeometric functions  \cite{OprisaWU}. Other occurrences of periods as physical quantities are string compactifications on Calabi--Yau manifolds.
According to Batyrev the period integrals of Calabi--Yau toric varieties are also governed by  GKZ systems. 
Therefore, the GKZ system is ubiquitous to functions describing physical effects in string theory as periods.

\section{Periods on $\bm{\Mo}$}
\label{sec:2}

The object of interest is the moduli space $\Mo$ of Riemann spheres (genus zero curves) of 
$N\geq 4$ ordered marked points  modulo the action of $PSL(2,\IC)$ on those points. The connected manifold $\Mo$  is described by the set of $N$--tuples of distinct points $(z_1,\ldots,z_N)$ modulo the action of $PSL(2,\IC)$ on those points. As a consequence with the choice 
\be\label{FIX}
z_1=0,\ \ \ z_{N-1}=1,\ \ \ z_N=\infty
\ee
there is a unique representative
\be\label{choice}
(z_1,\ldots,z_N)=(0,t_1,\ldots,t_{N-3},1,\infty)
\ee
of each equivalence class of $\Mo$
\be\label{MSP}
\Mo\simeq\{\ (t_1,\ldots,t_{N-3})\in \lf(\IP^1\backslash\{0,1,\infty\}\ri)^{N-3}\ |\ t_i\neq t_j\
 \mbox{for all}\ i\neq j\ \}\ ,
\ee
and the dimension of $\Mo(\IC)$ is $N-3$.
On the other hand, the real part of \req{MSP} describing the space of points
\be\label{StringDisk}
\Mo(\IR):=\{(0,t_1,\ldots,t_{N-3},1,\infty)\ |\ t_i\in \IR\}
\ee
is not connected. Up to dihedral permutation each of its $\h(N-1)!$ connected components  
(open cells $\gamma$) 
\be\label{cell}
\gamma=(z_1,z_2,\ldots,z_N)
\ee
is  completely described by the (real) ordering of the $N$ marked points
\be\label{ordering}
z_1<z_2<\ldots<z_N\ ,
\ee
with:
\be\label{subject} 
\bigcup_{i=1}^N \{z_i\}=\{0,t_1,\ldots,t_{N-3},1,\infty\}\ .
\ee
In the compactification $\ov\Mc_{0,N}(\IR)$ the components $\gamma$ become closed cells.
Each cell corresponds to a triangulation of a regular polygon with $N$ sides. The number of triangulations is given by $C_{N-2}=\fc{2^{N-2}(2N-5)!!}{(N-1)!}$ (with $C_N$ the Catalan number). In total an underlying $K_{N-1}$ associahedron (Stasheff polytope) can naturally be associated with  each vertex describing one triangulation \cite{BrownQJA}.
The standard cell of $\Mo$ is denoted by $\delta$ and given by the set of real marked points 
$(z_1,z_2,\ldots,z_N)=(0,t_1,t_2,\ldots,t_{N-3},1,\infty)$ on $\Mo$ subject to the (canonical) ordering
\req{ordering}, i.e.:
\be\label{standard}
\delta=\{\ t_l\in\IR\ |\ 0<t_1<t_2<\ldots<t_{N-3}<1\ \}\ .
\ee

A period on $\Mo$ is defined to be a convergent integral \cite{GM}
\be\label{PERIOD}
\int_\delta\omega
\ee
over the standard cell \req{standard} in $\Mo(\IR)$ and $\omega\in H^{N-3}(\Mo)$ a regular algebraic $(N-3)$--form, which converges on $\delta$ and has no poles along $\ov\delta$.
Every period on $\Mo$ is a $\IQ$--linear combination of MZVs \cite{BrownQJA}. Furthermore, every MZV
can be written as \req{PERIOD}.

To each cell $\gamma$ a unique $(N-3)$--form  can be associated \cite{BCS}
\be\label{cellf}
\omega_\gamma=\prod_{i=2}^N (z_i-z_{i-1})^{-1}\ dt_1\wedge\ldots\wedge dt_{N-3}\ ,
\ee
subject to \req{subject} with $z_l=\infty$ dismissed in the product. 
The form \req{cellf} is unique up to scalar multiplication, holomorphic on the interior of $\gamma$ and has  simple poles on the boundary of that cell. 
To a cell \req{ordering} in $\Mo(\IR)$  modulo rotations an oriented $N$--gon ($N$--sided polygons) may be associated by labelling clockwise its sides with the marked points $(z_1,z_2,\ldots,z_N)$.
E.g.~according to \req{ordering} the polygon with the cyclically labelled sides 
$\gamma=(0,1,t_1,t_3,\infty,t_2)$  is identified with the cell
$0<1<t_1<t_3<\infty<t_2$ in ${\cal M}_{0,6}(\IR)$ and the corresponding cell form is:
$$\omega_\gamma=\pm\fc{dt_1dt_2dt_3}{(-t_2)\ (t_3-t_1)\ (t_1-1)}\ .$$

The cell form \req{cellf} refers to the ordering \req{ordering}. A cyclic structure $\gamma$ 
corresponds to the cyclic ordering $(\gamma(1),\gamma(2),\ldots, \gamma(N))$ of
the elements $\{1,2,\ldots,N\}$ and refers to the standard $N$--gon $(1,2,\ldots,N)$ modulo rotations.
There is a unique ordering $\sigma$ of the $N$ marked points \req{choice} as 
\be\label{Ordering}
z_{\sigma(1)}<z_{\sigma(2)}<\ldots<z_{\sigma(N)}\ .
\ee
with $\sigma(N)=N$ and compatible with the cyclic structure $\gamma$. 
The cell--form corresponding to $\gamma$ is  defined  as \cite{BCS}
\be\label{Cell}
\omega_\gamma=\prod\limits_{i=2}^{N-1}\lf(z_{\sigma(i)}-z_{\sigma(i-1)}\ri)^{-1}dt_1\wedge\ldots\wedge dt_{N-3}\ .
\ee
E.g.~for the cyclic structure $(2,5,1,6,4,3)$ the unique  ordering $\sigma$ compatible with the latter 
and with  $\sigma(6)=6$ is the ordering $(4,3,2,5,1,6)$, i.e.~$\gamma=(t_3,t_2,t_1,1,0,\infty)$.

\begin{figure}[h]
\centering
\sidecaption
\includegraphics[width=0.35\textwidth]{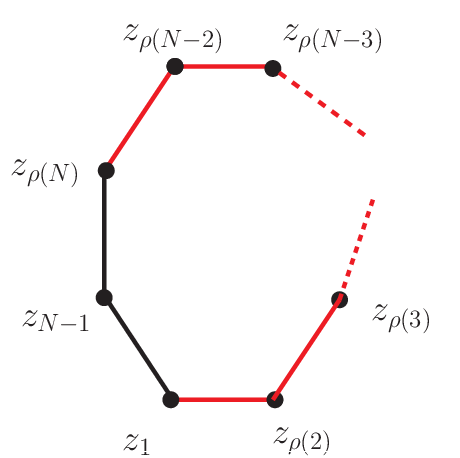}
\caption{$N$--gon describing the $01$ cyclic structure $\gamma=(0,1,\rho)$.}
\label{polygon}
\end{figure}

In the following, we consider orderings \req{ordering} ($01$ cyclic structure $\gamma$) 
of the set 
$\bigcup_{i=1}^N \{z_i\}=\{0,t_1,\ldots,t_{N-3},1,\infty\}$ with the elements $z_1=0$ and $z_{N-1}=1$ being
consecutive, i.e.~$\gamma=(0,1,\rho)$ with $\rho\in S_{N-2}$ some ordering of the $N-2$ 
points $\{t_1,\ldots,t_{N-3},\infty\}$. The corresponding  cell--function is given by
\be\label{cellb}
\omega_\rho=z_{\rho(2)}^{-1}\ \prod\limits_{i=3}^{N-2}\lf(z_{\rho(i)}-z_{\rho(i-1)}\ri)^{-1}dt_1\wedge\ldots\wedge dt_{N-3}\ \ \ ,\ \ \ \rho\in S_{N-2}\ ,
\ee
it is called $01$ cell--function \cite{BCS} and its associated $N$--gon, in which the edge referring to $0$ appears next to that referring to $1$, is depicted in Fig.~\ref{polygon}.

\noindent
The $(N-2)!$ $01$ cell--functions \req{cellb} generate the top--dimensional cohomology group $H^{N-3}(\Mo)$ of 
$\Mo$ by constituting a basis of $H^{N-3}(\Mo,\IQ)$, i.e.~\cite{BCS}:
\be
\dim H^{N-3}(\Mo,\IQ)=(N-2)!\ .
\ee
As a consequence the cohomology group $H^{N-3}(\Mo)$ is canonically isomorphic to the subspace of polygons 
having the vertex (edge) $0$ adjacent to edge $1$ \cite{BCS}.

Generically, in terms of cells a period \req{PERIOD} on $\Mo$ may be defined as the integral \cite{BCS}
\be\label{PERIODD}
\int_\beta\omega_\gamma
\ee
over the cell $\beta$ in $\Mo(\IR)$ and the cell--form $\omega_\gamma$ with the pair $(\beta,\gamma)$ referring to some polygon pair. Therefore, generically the cell--forms \req{cellf} integrated over cells \req{cell} give rise to periods on $\Mo$,
which are $\IQ$--linear combinations  of MZV.
By changing variables the period integral \req{PERIODD} can be brought into an integral over the standard cell $\delta$ parameterized in \req{standard}.
To obtain a convergent integral \req{PERIODD} in \cite{BCS} certain linear combinations of $01$ cell--forms \req{cellb} (called insertion forms) have been constructed with the properties of having no poles along the boundary of the standard cell $\delta$ and converging on the closure $\ov\delta$.
E.g.~in the case of $\Mc_{0,5}$ 
the cell--form $\omega_\gamma$ corresponding to the cell $\gamma=(0,1,t_1,\infty,t_2)$
can be integrated over the compact standard cell $\ov \delta$ defined in \req{standard}
\be
\int_{\ov\delta}\omega_\gamma=\int_{0\leq t_1\leq t_2\leq 1}\fc{dt_1dt_2}{(1-t_1)\ t_2}=\z_2\ ,
\ee
with the period $\zeta_2$ following from the general definition for the Riemann zeta function:
\be\label{Riemann}
\zeta_a=\sum_{k=1}^\infty k^{-a}\ \ \ ,\ \ \ a\in\IN,\ a\geq 2\ .
\ee

\section{Volume form and period matrix on $\bm{\Mo}$}

For a regular algebraic $(N-3)$--form $\omega_\delta$ on $\Mo$ conditions exist for the integral \req{PERIOD}  over the standard cell $\delta$ to converge. 
The set of all regular $(N-3)$--forms can be written in terms of the canonical cyclically
invariant form  \cite{BrownQJA}:
\be\label{cellc}
\omega_\delta=\fc{dt_1\wedge \ldots\wedge dt_{N-3}}{t_2\ (t_3-t_1)\ (t_4-t_2)\cdot \ldots\cdot(t_{N-3}-t_{N-5})\ 
(1-t_{N-4})}\ .
\ee
(Up to multiplication by $\IQ^+$) this form is  the canonical volume form on $\Mo(\IR)$ without zeros or poles along the standard cell \req{standard}.
An algebraic volume form $\Omega$ on $\Mo(\IR)$ may be supplemented by the $PSL(2,\IC)$ invariant factor 
$\prod_{i<j}^{N-1} |z_i-z_j|^{s_{ij}}$ (subject to \req{FIX} and with
some conditions on the parameter $s_{ij}$, which turn into physical
conditions, cf. \req{Mandel}) as
\be\label{Omega}
\Omega=\fc{dt_1\wedge \ldots\wedge dt_{N-3}}{t_2\ (t_3-t_1)\ (t_4-t_2)\cdot \ldots\cdot(t_{N-3}-t_{N-5})\ 
(1-t_{N-4})}\ \lf(\prod_{i<j}^{N-1} |z_i-z_j|^{s_{ij}}\ri)\ ,
\ee
with $s_{ij}\in\IZ$. The form \req{Omega} gives rise to the family of periods $\int_{\ov\delta}\Omega$
of $\Mo$
\be\label{MZVPeriod}
I_\delta(a,b,c)=\int_{\ov\delta}dt_1\cdot\ldots\cdot dt_{N-3}\   
 \prod\limits_{i=1}^{N-3}t_i^{a_i}\ (1-t_i)^{b_i}\ \prod\limits_{1\leq i<j\leq N-3}(t_i-t_j)^{c_{ij}}\ ,
\ee
for suitable choices of integers $a_i,b_i,c_{ij}\in\IZ$ such that the integral converges. The latter refers to the  compactified standard cell $\delta$ defined in \req{standard}.
It has been shown by Brown and Terasoma, that integrals of the form 
\req{MZVPeriod} yield linear combinations of MZVs with rational coefficients.
In cubical coordinates $x_1=\tfrac{t_1}{t_2},\ x_2=\tfrac{t_2}{t_3},\ldots,x_{N-4}=\tfrac{t_{N-4}}{t_{N-3}},\ x_{N-3}=t_{N-3}$ parameterizing  the integration region \req{standard} as
$t_k=\prod_{l=k}^{N-3}x_l,\ k=1,\ldots,N-3$ with $0<x_i<1$, the integral \req{MZVPeriod} becomes
\be\label{family}
I_\delta(a',b',c')=\lf(\prod_{i=1}^{N-3} \ \int^1_0 d x_i\ri) \ 
\prod_{j=1}^{N-3} \, x_j^{a'_j} \  (1-x_j)^{b'_j}\ \prod_{l=j+1}^{N-3} \ \left( \ 1 \ - \ \prod_{k=j}^l x_k \ \right)^{c_{jl}'}\ ,
\ee
with some integers $a_i',b_i',c_{ij}'\in\IZ$.

Moreover, the form \req{Omega}
 can be generalized to the family of real period integrals $\int_{\ov \delta}\Omega$ on $\ov \delta$, 
with $s_{ij}\in\IR$. Then, Taylor expanding \req{Omega} w.r.t.~$s_{ij}$ at integral points $s_{ij}\in\IZ^+$  yields coefficients representing period integrals of the form \req{MZVPeriod}. Similar observations have been made in \cite{OprisaWU,StiebergerTE} when computing $\ap$--expansions of string amplitudes which can be described by integrals of the type \req{family}. In this setup the additional $PSL(2,\IC)$ invariant factor 
$\prod_{i<j}^{N-1} |z_i-z_j|^{\ap s_{ij}}$ represents the so--called Koba--Nielsen factor with the
parameter $\ap$ being the inverse string tension and the kinematic invariants $s_{ij}$ specified in \req{Mandel}.

\begin{figure}[H]
\sidecaption
\includegraphics[width=0.35\textwidth]{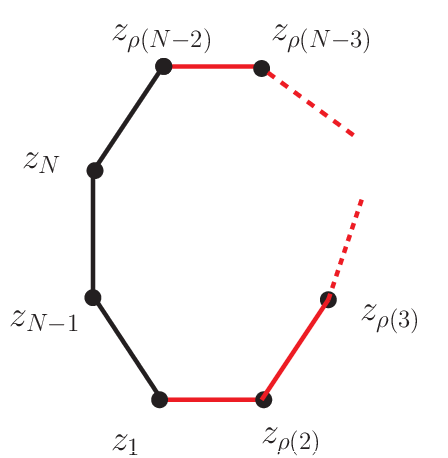}
\caption{$N$--gon describing the cyclic structure $\gamma=(0,1,\infty,\rho)$.}
\label{polygon2}
\end{figure}

Similarly to \req{Omega} in the following let us consider all the $(N-2)!$ $01$ cell--forms \req{cellb} supplemented by the $PSL(2,\IC)$ invariant factor $\prod_{i<j}^{N-1} |z_i-z_j|^{\ap s_{ij}}$ and integrated over the 
standard cell $\delta$ in $\Mo(\IR)$, i.e.: 
\be\label{cellc}
\int_{ \delta} \lf(\prod_{i<j}^{N-1} |z_i-z_j|^{\ap s_{ij}}\ri)\ \omega_\rho=\int_{\delta}
\fc{dt_1\wedge\ldots\wedge dt_{N-3}}{z_{\rho(2)}\ \prod\limits_{i=3}^{N-2}\lf(z_{\rho(i)}-z_{\rho(i-1)}\ri)}  \ \lf(\prod_{i<j}^{N-1} |z_i-z_j|^{\ap s_{ij}}\ri)\ \ \ ,\ \ \ \rho\in S_{N-2}\ .
\ee
Integration by part allows to express the $(N-2)!$ integrals \req{cellc} 
in terms of a basis of $(N-3)!$ integrals, i.e.:
\begin{svgraybox}
\be
\dim H^{N-3}(\Mo,\IR)=(N-3)!\ .
\ee
\end{svgraybox}
\noindent For a given  cell $\pi$ in $\Mo(\IR)$ we can choose the $01$ cell--form
$\omega_\gamma$ with $\gamma=(0,1,\infty,\rho),\ \rho\in S_{N-3}$ and
the following basis (subject to \req{FIX}) \cite{BroedelTTA}
\begin{eqnarray}
Z_\pi^\rho&:=&Z_\pi(1,\rho(2,3,\ldots,N-2),N,N-1) \label{ZZ}\\[3mm]
&= &\displaystyle{\int_{\pi} \lf(\prod_{i=2}^{N-2}dz_i\ri) \  \frac{\prod\limits_{i<j}^{N-1} |z_{ij}|^{\ap s_{ij}}}{  z_{1\rho(2)} z_{\rho(2),\rho(3)} \ldots z_{\rho(N-3),\rho(N-2)} } \ \ \ ,\ \ \ \pi,\rho\in S_{N-3}\ ,}\nonumber
\end{eqnarray}
with
\be\label{DIFF}
z_{ij}:=z_i-z_j\ ,
\ee
and $\rho$ describing some ordering of the $N-3$ 
points $\bigcup_{i=2}^{N-2}\{z_i\}=\{t_1,\ldots,t_{N-3}\}$ along the $N$--gon depicted in Fig.~\ref{polygon2}.

\noindent 
The iterated integrals \req{ZZ} represent generalized Euler (Selberg) integral and integrate to
multiple Gaussian hypergeometric functions \cite{OprisaWU}. Furthermore, the integrals \req{ZZ} can also be systematized within the framework of  Aomoto-Gelfand hypergeometric functions or GKZ structures~\cite{GKZ}.

The integrals \req{ZZ}
can be Taylor  expanded w.r.t.~$\ap$ around the  point $\ap=0$, e.g.:
\bea
&&\ds{\int\limits_{\delta}\lf(\prod_{l=2}^3dz_l\ri)\ \fc{\prod\limits_{i<j}^4|z_{ij}|^{\ap s_{ij}}}{z_{12}\ z_{23}\ z_{41}}}\\[7mm]
&=&\ds{\ap^{-2}\ \lf(\fc{1}{s_{12}s_{45}}+\fc{1}{s_{23}s_{45}}\ri)+\z_2\ \lf(1-\fc{s_{34}}{s_{12}}-\fc{s_{12}}{s_{45}}-\fc{s_{23}}{s_{45}}-\fc{s_{51}}{s_{23}}\ri)+\Oc(\ap)\ .}
\eea
Techniques for computing $\ap$ expansions for the type of integrals \req{ZZ} have been exhibited in 
\cite{OprisaWU,StiebergerTE},  systematized in \cite{BroedelTTA}, and pursued in \cite{Puhlfuerst:2015gta}.
In fact, the lowest order contribution of \req{ZZ} in the  Taylor  expansion  around the  point $\ap=0$
is given by
\be\label{Z0}
\lf.Z\ri|_{\ap^{3-N}}=(-1)^{N-3}\ S^{-1}\ ,
\ee
with the kernel\footnote{The matrix $S$ with entries $S_{\rho,\si}=S[\rho|\si]$ is defined as a $(N-3)!\times (N-3)!$ matrix with its rows and columns corresponding to the orderings $\rho \equiv \{\rho(2),\ldots,\rho(N-2)\}$ and
$\si \equiv \{\si(2),\ldots,\si(N-2)\}$, respectively. The matrix $S$ is symmetric, i.e.~$S^t=S$.}   \cite{KawaiXQ,BernSV,Bohr}
\be\label{kernel}
S[\rho|\si]:=S[\, \rho(2,\ldots,N-2) \, | \, \si(2,\ldots,N-2) \, ] = \ap^{N-3}\ \prod_{j=2}^{N-2} \Big( \, s_{1,j_\rho} \ + \ \sum_{k=2}^{j-1} \theta(j_\rho,k_\rho) \, s_{j_\rho,k_\rho} \, \Big)\ ,
\ee
with $j_\rho=\rho(j)$ and  $\theta(j_\rho,k_\rho)=1$
if the ordering of the legs $j_\rho,k_\rho$ is the same in both orderings
$\rho(2,\ldots,N-2)$ and $\si(2,\ldots,N-2)$, and zero otherwise. The matrix elements $S[\rho|\si]$
are polynomials of the order $N-3$ in the parameters \req{Mandel}.

A natural object to define is the $(N-3)!\times (N-3)!$--matrix
\be\label{Period}
F_{\pi\sigma}=(-1)^{N-3}\ \sum_{\rho\in S_{N-3}}Z_\pi(\rho)\ S[\rho|\sigma]\ ,
\ee
which according to \req{Z0} satisfies:
\be\label{FFT}
\lf.F\ri|_{\ap^{3-N}}=1\ .
\ee
The matrix $F$ has rank 
\begin{svgraybox}
\be
{\rm rk}(F)=(N-3)!
\ee
\end{svgraybox}
\noindent and represents the period matrix of $\Mo$ \cite{Sasha}.

In \cite{SS} it has been observed, that $F$ can be written in the following way\footnote{The ordering colons $:\ldots :$ are defined such that matrices with larger subscript multiply matrices with smaller subscript from the left, i.e.~$: \, M_{i} \ M_{j} \, : =  \lf\{
\begin{array}{lcl}M_{i} \ M_j\ , & i \geq j\ ,\\ 
       M_{j} \ M_i\ , &              i<j\ .\end{array}\ri.$
The generalization to iterated matrix products $: M_{i_1} M_{i_2} \ldots M_{i_p}:$ is straightforward.}
\be\label{Observation}
F=P\ Q\ :\exp\lf\{\sum\limits_{n\geq1}\zeta_{2n+1}\ M_{2n+1}\ri\}: \ ,
\ee
with the Riemann zeta--functions \req{Riemann}.
This decomposition is guided by its organization w.r.t.~multiple zeta values (MZVs) 
$\zeta_{n_1,\ldots,n_r}$ as
\be
\ba{lcl}
M_{2n+1}&=&\ds{\lf.F\ri|_{\z_{2n+1}}\ ,}\\[3mm]
P_{2n}&=&\ds{\lf.F\ri|_{\z_{2}^n}\ ,}
\ea\label{PM}
\ee
with:
\begin{eqnarray}
P&=&1+\sum_{n\geq 1}\z_2^n\ P_{2n}\ ,\label{PP}\\
Q&=&\displaystyle{1+\sum_{n\geq 8}Q_n=1+\fc{1}{5}\ \zeta_{3,5}\ [M_5,M_3]+\lf\{\ \fc{3}{14}\ \zeta_5^2+\fc{1}{14}\ \zeta_{3,7}\ \ri\}\ [M_7,M_3]}\nonumber\\
&+&\displaystyle{\lf\{\ 9\ \zeta_2\ \zeta_9+\fc{6}{25}\ \zeta_2^2\ \zeta_7-\fc{4}{35}\ \zeta_2^3\ \zeta_5+\fc{1}{5}\ \zeta_{3,3,5}\ \ri\}\ [M_3,[M_5,M_3]]+\ldots\ .}\label{QQQ}
\end{eqnarray}
MZVs are generalizations of single zeta functions \req{Riemann}
\be\label{MZV}
\zeta_{n_1,\ldots,n_r}:=\zeta(n_1,\ldots,n_r)=
\sum\limits_{0<k_1<\ldots<k_r}\ \prod\limits_{l=1}^r k_l^{-n_l}\ \ \ ,\ \ \ n_l\in\IN^+\ ,\ n_r\geq2\ ,
\ee
with $r$ specifying its depth and $w=\sum_{l=1}^rn_l$ denoting
its weight.
Hence, all the information is kept in the matrices $P$ and $M$ and the particular form of $Q$.
The entries of the matrices $M_{2n+1}$ are polynomials in $s_{ij}$ of degree $2n+1$ 
(and hence of the order $\ap^{2n+1}$), while the entries of the matrices $P_{2n}$ are polynomials in $s_{ij}$ of degree $2n$  (and hence of the order $\ap^{2n}$).
E.g.~for $N=5$ we have
\be
P=\ap^2\  \lf(\begin{array}{lcl}
-s_{34} s_{45}+s_{12}\ (s_{34}-s_{51})&s_{13}\ s_{24}\\
s_{12}\ s_{34}&(s_{12}+s_{23})\ (s_{23}+s_{34})-s_{45} s_{51}
\end{array}\ri)\ ,
\ee
and 
\be
M_3=\ap^3\ \lf(\begin{array}{lcl}
m_{11}&m_{12}\\ 
m_{21}&m_{22}
\end{array}\ri)\ ,
\ee
with:
\bea
m_{11}&=&s_{34}\ [\ -s_{12}\ (s_{12}+2 s_{23}+s_{34})+s_{34} s_{45}+s_{45}^2\ ]+s_{12} s_{51}\ (s_{12} +s_{51})\ ,\\[1mm]
m_{12}&=&-s_{13}\ s_{24}\ (s_{12}+s_{23}+s_{34}+s_{45}+s_{51})\ , \\[1mm] 
m_{21}&=&s_{12}\ s_{34}\ [\ s_{12}+s_{23}+s_{34}-2\ (s_{45}+s_{51})\ ]\ ,\\[1mm]
m_{22}&=&(s_{23}+s_{34})\ [\ (s_{12}+s_{23}) (s_{12}+s_{34})-2\ s_{12} s_{45}\ ]\\[1mm]
&&-[\ 2 s_{12} s_{34}-s_{45}^2+2 s_{23}\ (s_{34}+s_{45})\ ] 
s_{51}+s_{45} s_{51}^2\ .
\eea
As we shall see in section \ref{MotivicF} the form \req{Observation} is bolstered by the algebraic structure of motivic MZVs. The form \req{Observation} exactly appears in F. Browns decomposition of motivic MZVs \cite{Brown}.
In section \ref{Physics} we shall demonstrate, that the period matrix $F$ has also a physical meaning
describing scattering amplitudes of open and closed strings.

\section{Motivic and single--valued multiple zeta values}

MZVs \req{MZV} can be represented as period integrals. 
With the iterated integrals of the following form
\be
I_\gamma(a_0;a_1,\ldots,a_n;a_{n+1})=\int_{\Delta_n,\gamma} \fc{dz_1}{z_1-a_1}\ldots \fc{dz_n}{z_n-a_n}\ ,
\ee
with $\gamma$ a path in $M=\IC\slash \{a_1,\ldots,a_n\}$ with endpoints $\gamma(0)=a_0\in M,\ \gamma(1)=a_{n+1}\in M$ and $\Delta_{n,\gamma}$ a simplex  consisting of all ordered $n$--tuples of points $(z_1,\ldots,z_n)$ on $\gamma$
and for the map 
\be
\rho(n_1,\ldots,n_r)=10^{n_1-1}\ldots 10^{n_r-1}\ ,
\ee
with $n_r\geq 2$ Kontsevich observed that
\begin{eqnarray}
\zeta_{n_1,\ldots,n_r}&=&(-1)^r\ I_\gamma(0;\rho(n_1\ldots n_r);1)\nonumber\\[3mm]
&=&\ds{(-1)^r\ \int\limits_{0\leq t_1\leq\ldots\leq t_n\leq 1} \fc{dt_1}{t_1-a_1}\ldots \fc{dt_n}{t_n-a_n}\ ,}\label{integral}
\end{eqnarray}
with the sequence of numbers $(a_1,\ldots,a_n)$ given by $(1,0^{n_1-1},\ldots, 1,0^{n_r-1})$.
Note, that the integral \req{integral} defines a period.
Furthermore, the numbers \req{MZV} arise as coefficients of the 
Drinfeld associator $ Z(e_0,e_1)$ \cite{Drinii}. The latter is a function in terms of the generators $e_0$ and $e_1$ of a free Lie algebra $g$ and is given by the non--commutative generating series 
of (shuffle--regularized) MZVs \cite{LeMurakami}
\be
Z(e_0,e_1)=\sum_{w\in\{e_0,e_1\}^\times}\z(w)\ w\ ,
\label{drinfeld}
\ee
with the symbol $w\in\{e_0,e_1\}^\times$ denoting a non--commutative word 
$w_1w_2\ldots$ in the letters $w_i\in\{e_0,e_1\}$. 
Furthermore, we have the shuffle product
$\z(w_1)\z(w_2)=\z(w_1\shuffle w_2)$ and $\z(e_0)=0=\z(e_1)$ and $\z(e_1e_0^{n_1-1}\ldots e_1e_0^{n_r-1})=\z_{n_1,\ldots,n_r}$. 
Explicitly,  \req{drinfeld} becomes:
\begin{eqnarray}
Z(e_0,e_1) &=&\sum_{w\in\{e_0,e_1\}^\times}\z(w)\ w=1+\z_2\ [e_0,e_1]+\z_3\ (\ [e_0,[e_0,e_1]]-[e_1,[e_0,e_1]\ )\nonumber\\
&+&\z_4\ \Big(\ [e_0,[e_0,[e_0,e_1]]]-\fc{1}{4}\ [e_1,[e_0,[e_0,e_1]]]+[e_1,[e_1,[e_0,e_1]]]
+\fc{5}{4}\ [e_0,e_1]^2\ \Big)\nonumber\\
&+&\z_2\ \z_3\ \Big(\ ( [e_0,[e_0,e_1]]-[e_1,[e_0,e_1] )\ [e_0,e_1]+[e_0,[e_1,[e_0,[e_0,e_1]]]]\cr
&-&[e_0,[e_1,[e_1,[e_0,e_1]]]]\ \Big)
+\z_5\ \Big(\ [e_0,[e_0,[e_0,[e_0,e_1]]]]\nonumber\\
&-&\h\ [e_0,[e_0,[e_1,[e_0,e_1]]]]-\fc{3}{2}\ [e_1,[e_0,[e_0,[e_0,e_1]]]]
+(e_0\leftrightarrow e_1)\ \Big)+\ldots\ .\label{Drinfeld}
\end{eqnarray}
The set of integral linear combinations of MZVs \req{MZV} is a ring, since the product of any two values can be expressed by a
(positive) integer linear combination of the other MZVs \cite{Zagier}.
There are many relations over $\IQ$ among MZVs.
We define the (commutative) $\IQ$--algebra $\Zc$ spanned by all MZVs over $\IQ$. The latter is the 
(conjecturally direct) sum over the  
$\IQ$--vector spaces $\Zc_N$ spanned by the set of MZVs \req{MZV} of total weight $w=N$, with $n_r\geq2$,
 i.e.~$\Zc=\bigoplus_{k\geq 0}\Zc_k$. 
For a given weight $w\in\IN$ the dimension $\dim_\IQ(\Zc_N)$ of the space $\Zc_N$  is conjecturally given
by $\dim_\IQ(\Zc_N)=d_N$, with $d_N=d_{N-2}+d_{N-3},\ N\geq 3$ and $d_0=1,\ d_1=0,\ d_2=1$ \cite{Zagier}. Starting at weight $w=8$ MZVs of depth greater than one $r>1$ appear
in the basis.
By applying stuffle, shuffle, doubling, generalized doubling relations and duality it is possible to reduce the MZVs of a given weight to a minimal set.
Strictly speaking this is explicitly proven only up to weight $26$ \cite{DataMine}.
For $D_{w,r}$ being the number of independent MZVs at weight $w>2$ and depth $r$, which cannot be 
reduced to primitive MZVs of smaller depth and their products,  
it is believed, that $D_{8,2}=1, 
D_{10,2}=1,\ D_{11,3}=1,\ D_{12,2}=1$ and $D_{12,4}=1$~\cite{Broadii}. 
For $Z=\fc{\Zc_{>0}}{\Zc_{>0}\cdot\Zc_{>0}}$ with $\Zc_{>0}=\oplus_{w>0}\Zc_w$ the graded space of irreducible MZVs we have:
$dim(Z_w)\equiv\sum_r D_{w,r}=1,0,1,0,1,1,1,1,2,2,3,3,4,5$ for $w=3,\ldots,16$, 
respectively \cite{Broadii,DataMine}.

An important question is how to decompose a MZV of a certain weight $w$
in terms of a given basis of the same weight $w$.
E.g.~for the decomposition
\be
\z_{4,3,3}=\fc{4336}{1925}\ \z_2^5+\fc{1}{5}\ \z_2^2\z_3^2+10\ \z_2\z_3\z_5-\fc{49}{2}\ \z_5^2-18\ \z_3\z_7-4\ \z_2\z_{3,5}+\z_{3,7}
\ee
we wish to find a method to determine the rational coefficients.
Clearly, this question cannot be answered within the space of MZV $\Zc$ as we do not know
how to construct a basis of MZVs for any weight.
Eventually, we seek to answer the above question within the space $\Hc$ of motivic MZVs
with the latter serving as some auxiliary objects for which we assume certain properties \cite{Brown}.
For this purpose the actual MZVs \req{MZV} are replaced by symbols (or motivic MZVs), 
which are elements of a certain algebra. We lift the ordinary MZVs $\z$
to their motivic versions\footnote{In \cite{Goncharov,GoncharovBrown} motivic MZVs $\zeta^{\frak m}$ are defined as elements of a certain graded algebra $\Hc$ equipped with a period homomorphism \req{per}.} $\z^{\frak m}$ with the surjective projection (period map) \cite{Goncharov,GoncharovBrown}:
\be\label{per}
\per: \z^{\frak m}\longrightarrow \z\ .
\ee
Furthermore, the standard relations  among MZV (like shuffle and stuffle relations) are supposed to hold for the motivic MZVs $\z^{\frak m}$.
In particular, $\Hc$ is a graded Hopf algebra\footnote{A Hopf algebra
is an algebra $\Ac$ with multiplication $\mu: \Ac\otimes\Ac\ra\Ac$, i.e.~$\mu(x_1\otimes x_2)=x_1\cdot x_2$ and associativity. At the same time it is also a coalgebra with coproduct $\Delta: \Ac\ra\Ac\otimes\Ac$  and coassociativity such that the product and coproduct are compatible: $\Delta(x_1\cdot x_2)=\Delta(x_1)\cdot\Delta(x_2)$, with $x_1,x_2\in\Ac$.} $\Hc$ with a coproduct $\Delta$, i.e.~
\be\label{gradedalgebra}
\Hc=\bigoplus_{n\geq 0}\Hc_n\ ,
\ee
 and for each weight $n$ the Zagier conjecture is assumed to be true, i.e.~
$\dim_\IQ(\Hc_n)=d_n$.
To explicitly describe the structure of the space $\Hc$ one  introduces the (trivial)
 algebra--comodule:
\be\label{introU}
\Uc=\IQ\vev{f_3,f_5,\ldots}\ \otimes_\IQ\ \IQ[f_2]\ .
\ee
The multiplication on 
\be\label{Up}
\Uc'=\Uc\big\slash f_2\Uc=\IQ\vev{f_3,f_5,\ldots}
\ee 
is given by the shuffle product $\shuffle$
\be\label{schuffle}
f_{i_1}\ldots f_{i_r}\shuffle f_{i_{r+1}}\ldots f_{i_{r+s}}=\sum_{\si\in\Si(r,s)}
f_{i_{\si(1)}}\ldots f_{i_{\si(r+s)}}\ ,
\ee
$\Si(r,s)=\{\si\in\Si(r+s)\ | \ \si^{-1}(1)<\ldots<\si^{-1}(r) \cap
\si^{-1}(r+1)<\ldots<\si^{-1}(r+s) \}$. The Hopf--algebra  $\Uc'$ is 
isomorphic to the space of non--commutative polynomials in $f_{2i+1}$.
The element $f_2$ commutes with all $f_{2r+1}$. Again, there is a grading $\Uc_k$ on $\Uc$,
with $\dim(\Uc_k)=d_k$.
Then, there exists a morphism $\phi$ of graded algebra--comodules
\be\label{morphism}
\phi:\ \Hc\lra\Uc\ ,
\ee
normalized\footnote{Note, that there is no canonical choice of $\phi$ and the latter depends  on the choice
of motivic generators of~$\Hc$.} by:
\be\label{morphismres}
\phi\big(\zeta^{\frak m}_n\big)=f_n\ \ \ ,\ \ \ n\geq 2\ .
\ee
Furthermore, \req{morphism} respects the shuffle multiplication rule \req{schuffle}:
\be\label{ruleshuffle}
\phi(x_1x_2)=\phi(x_1)\shuffle \phi(x_2)\ \ \ ,\ \ \ x_1,x_2\in\Hc\ .
\ee
The map \req{morphism} is defined recursively from lower weight and sends every motivic MZV $\xi\in\Hc_{N+1}$ of weight $N+1$
to a non--commutative polynomial in the $f_i$. The latter is given as series expansion up to weight $N+1$
w.r.t.~the basis $\{f_{2r+1}\}$
\be
\phi(\xi)=c_{N+1}\ f_{N+1}+\sum_{3\leq 2r+1\leq N}f_{2r+1}\ \xi_{2r+1}\in\Uc_{N+1}\ ,
\ee
with the coefficients $\xi_{2r+1}\in \Uc_{N-2r}$ being of smaller weight than $\xi$
and computed from the coproduct as follows. The derivation $D_r:\Hc_m\ra \Ac_r\otimes \Hc_{m-r}$, with
$\Ac=\Hc/\z_2\Hc$  takes only a subset of the full coproduct, namely the weight $(r,m-r)$--part. Hence,
$D_{2r+1}\xi$ gives rise to a weight $(2r+1,N-2r)$--part $x_{2r+1}\otimes y_{N-2r}\in \Ac_{2r+1}\otimes \Hc_{N-2r}$ and $\xi_{2r+1}:=c_{2r+1}^\phi(x_{2r+1})\cdot \phi(y_{N-2r})$ 
The operator $c_{2r+1}^\phi(x_{2r+1})$, with $x_{2r+1}\in\Ac_{2r+1}$ determines the rational coefficient of $f_{2r+1}$ in the monomial $\phi(x_{2r+1})\in\Uc_{2r+1}$.
Note, that the right hand side of $\xi_{2r+1}$ only involves elements  from $\Hc_{\leq N}$
for which $\phi$ has already been determined.
On the other hand, the coefficient $c_{N+1}$ cannot be determined by this method unless we specify\footnote{The choice of $\phi$ describes for each weight $2r+1$ the motivic derivation operators $\p_{2r+1}^\phi$  acting on the space of motivic MZVs $\p_{2r+1}^\phi:\Hc\ra\Hc$ \cite{Brown} as:
\be\label{motder}
\p_{2r+1}^\phi=(c_{2r+1}^\phi\otimes id) \circ D_{2r+1}\ ,
\ee
with  the coefficient function $c_{2r+1}^\phi$.} 
a basis $B$ and compute $\phi$ for this basis giving rise to the basis dependent map $\phi^B$. 
E.g.~for the basis $B=\{\z_2^{\frak m}\z_3^{\frak m},\z_5^{\frak m}\}$ we have
$\phi^B(\z_2^{\frak m}\z_3^{\frak m})=f_2f_3$ and $\phi^B(\z_5^{\frak m})=f_5$, while
$\phi^B(\z^{\frak m}_{2,3})=3f_3f_2+cf_5$ with $c$ undetermined.

To illustrate the procedure for computing the map \req{morphism} and determining the decomposition
let us consider the case of weight $10$.
First, we introduce a basis of motivic MZVs
\be
B_{10}=\{\ \MZ{3,7},\ \MZ{3}\ \MZ{7},\ (\MZ{5})^2,\ \MZ{3,5}\ \MZ{2},\ 
\MZ{3}\ \MZ{5}\ \MZ{2},\ (\MZ{3})^2\ (\MZ{2})^2,\ (\MZ{2})^5\ \}\ ,
\ee
with $\dim(B_{10})=d_{10}$.
Then for each basis element we compute \req{morphism}:
\be\label{Ex10}
\ba{lcl}
\ds{\phi^B\lf(\MZ{3,7}\ri)}&=&\ds{-14\ f_7f_3-6\ f_5f_5,\ \ \ \phi^B\lf(\MZ{3}\MZ{7}\ri)=
f_3\shuffle f_7\ ,}\\[3mm]
\ds{\phi^B\lf((\MZ{5})^2\ri)}&=&\ds{f_5\shuffle f_5,\ \ \ 
\phi^B\lf(\MZ{3,5}\MZ{2}\ri)=-5\ f_5f_3f_2\ ,}\\[3mm]
\ds{\phi^B\lf(\MZ{3}\MZ{5}\MZ{2}\ri)}&=&\ds{f_3\shuffle f_5 f_2,\ \ \ 
\phi^B\lf((\MZ{3})^2(\MZ{2})^2\ri)=f_3\shuffle f_3 f_2^2\ ,}\\[3mm]
\ds{\phi^B\lf((\MZ{2})^5\ri)}&=&\ds{f_2^5\ .}
\ea\
\ee

The above construction allows to assign a $\IQ$--linear combination of monomials to every element
$\zeta^{\frak m}_{n_1,\ldots,n_r}$. 
 The map \req{morphism} sends every motivic MZV of weight less or equal to $N$ to a non--commutative polynomial in the $f_i$'s. Inverting the map $\phi$  gives the decomposition of $\zeta^{\frak m}_{n_1,\ldots,n_r}$ w.r.t.~the basis $B_n$ of weight $n$, with $n=\sum_{l=1}^rn_l$.
We construct operators acting on $\phi(\xi)\in \Uc$ to detect elements 
in $\Uc$ and to decompose any motivic MZV $\xi$ into a candidate basis~$B$.
The derivation operators $\p_{2n+1}:\Uc\ra\Uc$  are defined as \cite{Brown}:
\be\label{defder}
\p_{2n+1}(f_{i_1}\ldots f_{i_r})=\lf\{\begin{array}{lcl}
                                        f_{i_2}\ldots f_{i_r}, & i_1=2n+1\ ,\\
                                        0\ , & \rm{otherwise}\ ,
                                        \end{array}\ri.
                                        \ee
with $\p_{2n+1} f_2=0$.
Furthermore, we have the product rule for the shuffle product:
\be\label{Leibniz}
\p_{2n+1}(a\shuffle b)=\p_{2n+1} a\shuffle b+a\shuffle \p_{2n+1} b\ \ \ ,\ \ \ a,b\in \Uc'\ .
\ee
Finally, $c_2^n$ takes the coefficient of $f_2^n$.
By first determining the map \req{morphism} for a given basis $B_n$ we then can construct 
the motivic decomposition operator $\xi_n$ such that it acts trivially on this basis.
This is established for the weight ten  basis \req{Ex10} in the following.

With the differential operator \req{defder} we may consider the following operator
\be
\ba{lcl}
\ds{\xi_{10}}&=&\ds{a_0\ (\MZ{2})^5+a_1\ (\MZ{2})^2\ (\MZ{3})^2+a_2\ \MZ{2}\ \MZ{3}\ \MZ{5}+
a_3\ (\MZ{5})^2}\\[3mm]
&+&\ds{a_4\ \MZ{2}\ \MZ{3,5}+a_5\ \MZ{3}\ \MZ{7}+a_6\ \MZ{3,7}}
\ea\label{Ex10a}
\ee
with the operators 
\be
\ba{lcl}
a_1&=&\h\ c_2^2\ \p_3^2,\ a_2=c_2\ \p_5\p_3,\ a_3=\h\ \p_5^2+
\fc{3}{14}\ [\p_7,\p_3]\\[3mm]
a_4&=&\fc{1}{5}\ c_2\ [\p_5,\p_3],\ a_5=\p_7\p_3,\ a_6=\fc{1}{14}\ [\p_7,\p_3]
\ea\label{Ex10b}
\ee
acting on $\phi^B(\xi_{10})$.
Clearly, for the basis \req{Ex10} we exactly verify \req{Ex10a} to a be a decomposition operator
acting trivially on the basis elements.

Let us now discuss  a special class of MZVs  \req{MZV} identified as single--valued MZVs (SVMZVs)
\be\label{trivial}
\SV(n_1,\ldots,n_r)\in\IR
\ee
originating from single--valued multiple polylogarithms (SVMPs) at unity \cite{BrownPoly}. 
The latter are generalization of the Bloch--Wigner dilogarithm:
\be
D(z)=\Im\lf\{{\cal L}i_2(z)+\ln|z|\ \ln(1-z)\ \ri\}\ .
\ee
Thus, e.g.: 
\be
\SV(2)=D(1)=0\ .
\ee 
SVMZVs represent a subset of the MZVs \req{MZV} and they  satisfy the same double shuffle and associator relations than the
usual MZVs  and many more relations~\cite{SVMZV}. SVMZVs have  recently  been studied by Brown in \cite{SVMZV} from a mathematical point of view. They have been identified as the coefficients in an infinite series expansion of the Deligne associator 
\cite{Deligne}\ in two non--commutative variables. The latter is defined through the equation \cite{SVMZV}
\be\label{deligne}
W(e_0,e_1)=Z(-e_0,-e_1')^{-1}\ Z(e_0,e_1)\ ,
\ee
with the Drinfeld associator \req{Drinfeld} and $e_1'=We_1W^{-1}$.
The equation \req{deligne} can systematically be worked out at each weight yielding \cite{StiebergerWEA}:
\be
\begin{array}{lcl}
W(e_0,e_1)&=&1+2\ \z_3\ ([e_0,[e_0,e_1]]-[e_1,[e_0,e_1])+2\ \z_5\ 
\Big([e_0,[e_0,[e_0,[e_0,e_1]]]] \\
&-&\h\ [e_0,[e_0,[e_1,[e_0,e_1]]]]-\fc{3}{2}\ [e_1,[e_0,[e_0,[e_0,e_1]]]]
+(e_0\leftrightarrow e_1)\Big)+\ldots\ .
\end{array}
\label{DeligneA}
\ee
Strictly speaking, the numbers \req{trivial} are established in the Hopf algebra \req{gradedalgebra} of motivic MZVs $\zeta^{\frak m}$. 
In analogy to the motivic version of the Drinfeld associator \req{Drinfeld}
\be\label{MotDrinfeld}
Z^{\frak m}(e_0,e_1)=\sum_{w\in\{e_0,e_1\}^\times}\z^{\frak m}(w)\ w
\ee
in Ref.~\cite{SVMZV} Brown has defined the motivic single--valued associator as a generating series
\be\label{motDeligne}
W^{\frak m}(e_0,e_1)=\sum_{w\in\{e_0,e_1\}^\times}\z_{\rm sv}^{\frak m}(w)\ w\ ,
\ee
whose period map \req{per} gives the Deligne associator \req{deligne}.
Hence, for the motivic MZVs there is a map from the motivic MZVs to SVMZVs furnished by  the following homomorphism 
\be\label{HOMO}
\sv: \Hc\ra\Hc^\sv\ ,
\ee
with:
\be\label{motivicSV}
\sv: \z^{\frak m}_{n_1,\ldots,n_r}\mapsto\SVM(n_1,\ldots,n_r)\ .
\ee
In the  algebra $\Hc$ the homomorphism \req{HOMO} together with 
\be\label{Withh}
\SVM(2)=0
\ee
can be constructed \cite{SVMZV}.
The motivic SVMZVs $\SVM(n_1,\ldots,n_r)$ generate the subalgebra $\Hc^\sv$
of the Hopf algebra $\Hc$ and satisfy all motivic relations between MZVs.

In practice, the map $\sv$ is constructed recursively
in the (trivial) algebra--comodule  \req{introU}  with the
first factor given by \req{Up} and generated by all non--commutative words in the letters 
$f_{2i+1}$. We have $\Hc\simeq\Uc$, in particular $\z^{\frak m}_{2i+1}\simeq f_{2i+1}$.  The homomorphism
\be\label{mapsv}
\sv:\Uc'\longrightarrow\Uc'\ ,
\ee
with
\be\label{Mapsv}
w\longmapsto \sum_{uv=w} u\shuffle \tilde v\ ,
\ee
and
\be\label{Withhh}
\sv(f_2)=0
\ee
maps  the algebra of non--commutative words $w\in\Uc$ to the smaller subalgebra $\Uc^\sv$, which describes  the space of SVMZVs \cite{SVMZV}.
In eq. \req{Mapsv} the word $\tilde v$ is the reversal of the word $v$ and $\shuffle$ is the shuffle product.
For more details we refer the reader to the original reference \cite{SVMZV} and subsequent
applications in \cite{StiebergerWEA}.
With \req{Mapsv} the image of $\sv$ can be computed very easily, e.g.: 
\be
\sv(f_{2i+1})=2f_{2i+1}\ .
\ee
Eventually, the period map \req{per} implies the  homomorphism
\be\label{mapSV}
\sv: \z_{n_1,\ldots,n_r}\mapsto\ \SV( n_1,\ldots,n_r )\ ,
\ee
and with \req{Mapsv} we find the following examples (cf.~Ref.~\cite{StiebergerWEA} for more examples):
\begin{eqnarray}
\sv(\z_2)&=&\SV(2)=0\ ,\\
\sv(\z_{2n+1})&=&\SV(2n+1)=2\ \zeta_{2n+1}\ ,\ \ \ n\geq 1\ ,\\
\sv(\z_{3,5})&=&-10\ \z_3\ \z_5\ \ \ ,\ \ \ \sv(\z_{3,7})=-28\ \z_3\ \z_7-12\ \z_5^2\ ,\\
\sv(\z_{3,3,5})&=&2\ \z_{3,3,5}-5\ \z_3^2\ \z_5+90\ \z_2\ \z_9+\fc{12}{5}\ \z_2^2\ \z_7-\fc{8}{7}\ \z_2^3\ \z_5^2\ ,\ldots\ .
\end{eqnarray}

\section{Motivic period matrix $\bm{F^{\frak m}}  $}
\label{MotivicF}

The motivic version $F^{\frak m}$ of the period matrix \req{Observation} is given by passing from the MZVs $\zeta\in\Zc$ to their motivic versions $\zeta^{\frak m}\in\Hc$ as
\be\label{Fmotivic}
F^{\frak m}=P^{\frak m}\ Q^{\frak m}\ :\exp\lf\{\sum_{n\geq 1}\zeta_{2n+1}^{\frak m}\ M_{2n+1}\ri\}:\ ,
\ee
with
\be
P^{\frak m}=\lf.P\ri|_{\zeta_2\ra \z_2^{\frak m}}\ \ \ ,\ \ \ Q^{\frak m}=\lf.Q\ri|_{\zeta_{n_1,\ldots,n_r}\ra\zeta^{\frak m}_{n_1,\ldots,n_r}}\ ,
\ee
and the matrices $P,M$ and $Q$ defined in \req{PM} and \req{QQQ}, respectively.
Extracting e.g.~the weight $w=10$ part of \req{Fmotivic}
\begin{eqnarray}
\lf.F^{\frak m}\ \ri|_{\zeta^{\frak m}_3\zeta^{\frak m}_7}&=&M_7\ M_3\ ,\nonumber\\[2mm]
\lf.F^{\frak m}\ \ri|_{\zeta^{\frak m}_{3,7}}&=&\fc{1}{14}\ \lf[M_7,M_3\ri]\ ,\nonumber\\[2mm]
\lf.F^{\frak m}\ \ri|_{(\zeta^{\frak m}_5)^2}&=&\h\ M_5^2+\fc{3}{14}\ \lf[M_7,M_3\ri]\ ,\nonumber\\[2mm]
\lf.F^{\frak m}\ \ri|_{\z^{\frak m}_2\zeta^{\frak m}_3\zeta^{\frak m}_5}&=&P_2\ M_5\ M_3\ ,\nonumber\\[2mm]
\lf.F^{\frak m}\ \ri|_{\z^{\frak m}_2\zeta^{\frak m}_{3,5}}&=&\fc{1}{5}\ P_2\ \ \lf[M_5,M_3\ri]\ ,\nonumber\\[2mm]
\lf.F^{\frak m}\ \ri|_{(\z^{\frak m}_2)^2(\zeta^{\frak m}_{3})^2}&=&\fc{1}{2}\ P_4\ \ M_3^2\ ,\nonumber\\[2mm]
\lf.F^{\frak m}\ \ri|_{(\z^{\frak m}_2)^5}&=&P_{10}\ ,
\end{eqnarray}
and comparing with the motivic decomposition operators \req{Ex10b} yields a striking 
exact match in the coefficients and commutator structures by identifying the motivic derivation operators with the matrices \req{PM} as:
\begin{eqnarray}
\p_{2n+1}&\simeq& M_{2n+1}\ \ \ ,\ \ \ n\geq1\ ,\nonumber\\[3mm]
c_2^k&\simeq& P_{2k}\ \ \ ,\ \ \ k\geq1\ .\label{ExactMatch}
\end{eqnarray}
This agreement has been shown to exist up to the weight~$w=16$ in \cite{SS} and extended through 
weight~$w=22$ in \cite{BroedelTTA}. 
Hence, at least up to the latter weight  the decomposition of motivic MZVs w.r.t.~to a basis of MZVs encapsulates the 
$\ap$--expansion of the motivic period matrix written in terms of the same basis elements \req{Fmotivic}.

In the following we shall demonstrate, that the isomorphism  \req{morphism} encapsulates all the relevant  information of the $\ap$--expansion of the 
motivic period matrix \req{Fmotivic} without further specifying the latter explicitly 
in terms of motivic MZVs $\z^{\frak m}$. In the sequel we shall apply the isomorphism $\phi$ to $F^{\frak m}$.
The action \req{morphism} of $\phi$ on the motivic MZVs is explained in the previous section. 
The first hint of a simplification under $\phi$ occurs by considering the weight $w\!=\!8$ contribution to 
$F^{\frak m}$, 
where the  commutator term $[M_5,M_3]$ from  $Q_8^{\frak m}$ together with the 
prefactor ${1\over 5}\zeta_{3,5}^{\frak m}$ conspires\footnote{Note the useful relation $\phi^B(Q_8^{\frak m})=f_5f_3\ [M_3,M_5]$ for $Q_8^{\frak m}={1\over 5}\zeta_{3,5}^{\frak m} \ [M_5,M_3]$.} into (with $\phi^B(\z_{3,5}^{\frak m})=-5f_5f_3$):
\be\label{testx}
\phi^B\lf(\ \zeta_3^{\frak m} \zeta_5^{\frak m} \ M_5 M_3  + Q_8^{\frak m}\ \ri)= f_3 f_5 \ M_5 M_3 \ + \ f_5 f_3 \ M_3 M_5\ .
\ee
The right hand side obviously treats the objects $f_3,M_3$ and $f_5,M_5$ in a democratic way.
The effect of the map $\phi$ is, that in the Hopf algebra ${\cal U}$, every non--commutative word 
of odd letters $f_{2k+1}$ multiplies the  associated 
reverse product of matrices $M_{2k+1}$. 
Powers $f_2^k$ of the commuting generator $f_2$ are accompanied by  $P_{2k}$, which multiplies all the 
 operators  $M_{2k+1}$  from the left. 
Most notably, in contrast to the representation in terms of motivic MZVs, 
the numerical factors become unity, i.e.~all the rational numbers in \req{QQQ} drop out. 
Our explicit results confirm, that the  beautiful structure
with the combination of 
operators $M_{i_p} \ldots M_{i_2} M_{i_1}$ accompanying the word $f_{i_1} f_{i_2} \ldots f_{i_p}$,
continues to hold through at least weight $w=16$.
To this end, we obtain the following striking and short form for the motivic period matrix $F^{\frak m}$ \cite{SS}:
\be\label{summtwo}
\phi^B(F^{\frak m}) =\left( \sum_{k=0}^{\infty} f_2^k\ P_{2k} \right)\ \lf( \sum_{p=0}^{\infty} \sum_{ i_1,\ldots, i_p \atop  \in 2 \IN^+ + 1}
f_{i_1} f_{i_2}\ldots f_{i_p}\  M_{i_p} \ldots M_{i_2} M_{i_1}\ \ri)\ .
\ee
In \req{summtwo} the sum over the combinations $ f_{i_1} f_{i_2}\ldots f_{i_p} M_{i_p} \ldots M_{i_2} M_{i_1}$ includes 
{\it all} possible 
non--commutative words $f_{i_1} f_{i_2} \ldots f_{i_p}$ with coefficients 
$M_{i_p} \ldots M_{i_2} M_{i_1}$ graded by their length $p$. 
Matrices $P_{2k}$ associated with the  powers $f_2^k$ always act by left multiplication. The commutative nature of $f_2$ w.r.t.~the odd 
generators $f_{2k+1}$ ties in with the fact that in the matrix ordering the  matrices $P_{2k}$ have the well--defined place left of all matrices $M_{2k+1}$.
Alternatively, we may write \req{summtwo} in terms of a geometric series:
\begin{svgraybox}
\be
\phi^B(F^{\frak m}) =\left( \sum_{k=0}^{\infty} f_2^k\ P_{2k} \right)\ \lf(1-\sum_{k=1}^\infty f_{2k+1}\ M_{2k+1}\ri)^{-1}\ .
\ee
\end{svgraybox}
\noindent 
Thus, under the map $\phi$ the motivic period matrix $F^{\frak m}$ takes a very simple structure $\phi^B(F^{\frak m})$ in terms of the Hopf--algebra.

After replacing in \req{summtwo} the matrices \req{PM} by the operators as in \req{ExactMatch}
the operator \req{summtwo} becomes the canonical element in $\Uc\otimes \Uc^\ast$, which maps
 any non--commutative word in $\Uc$ to itself. In this representation \req{summtwo}
 gives rise to a group like action on $\Uc$.
 Hence, the  
operators $\p_{2n+1}$ and $c_{2}^k$ are dual to the letters $f_{2n+1}$ and $f_2^k$ and 
 have the matrix representations $M_{2n+1}$ and $P_{2k}$, respectively. 
By mapping the motivic MZVs $\z^{\frak m}$ of the period matrix $F^{\frak m}$ to elements $\phi^B(\zeta^{\frak m})$ of the Hopf algebra $\Uc$ the map  $\phi$ endows $F^{\frak m}$ with its motivic structure:
it maps the latter into a
very short and intriguing form in terms of the non--commutative Hopf algebra $\Uc$.
In particular, the various relations among different MZVs become simple algebraic 
identities in the Hopf algebra $\Uc$. 
Moreover, in this representation  the final result \req{summtwo} for period matrix does not depend on the choice of a specific set\footnote{For instance instead of taking a basis containing
the depth one elements $\zeta^{\frak m}_{2n+1}$ 
one also could choose the set of Lyndon words in the Hoffman elements
$\zeta^{\frak m}_{n_1,\ldots,n_r}$, with $n_i=2,3$ and define the corresponding matrices \req{PM}.}
 of MZVs as basis elements.
In fact, this feature follows from the basis--independent statement in terms of the motivic coaction
 (subject to matrix multiplication) \cite{Drummond:2013vz}
\be
\Delta F^{\frak m}=F^a \otimes F^{\frak m}\ ,
\ee
with the superscripts $a$ and $m$ referring to the algebras $\Ac$ and $\Hc$, respectively.
Furthermore with  \cite{BSST}
\be
\p_{2n+1}F^{\frak m}=F^{\frak m}\ M_{2n+1}
\ee
one can explicitly prove \req{summtwo}.

It has been pointed out in \cite{Brown:2015qmm} that the simplification occurring in \req{summtwo}
can be interpreted as a compatibility between the motivic period matrix and the action of the Galois
group of periods.
Let us introduce the free graded Lie algebra $\Fc$ over $\IQ$, which is freely generated by the symbols 
$\tau_{2n+1}$ of degree $2n+1$. Ihara has studied this algebra 
to relate the Galois Lie algebra $\Gc$ of the Galois group $G$ to the more tractable object $\Fc$ \cite{YI1}.
The dimension $\dim(\Fc_m)$ can explicitly given by \cite{TS}
\be\label{NICE1}
\dim(\Fc_m)=\sum_{d|m}\fc{1}{d}\ \mu(d)\ \sum_{\lceil\fc{m}{3d}\rceil\leq n \leq \lfloor\fc{m-d}{2d}\rfloor}\fc{1}{n}\ \lf(n\atop \fc{m}{d}-2n\ri)\ ,
\ee
with the M\"obius function $\mu$. 

\break
\thispagestyle{empty}
\setcounter{table}{0}

\begin{table}[H]
\caption{Linearly independent elements in $\Lc_m$ and primitive MZVs for $m=1,\ldots,22$.}
\label{tabs}
\begin{tabular}{|c|c|c|c|}
\hline
$m$			
&$\dim(\Fc_m)$
&{\rm linearly independent elements at} $\ap^{\frak m}$
&{\rm irreducible MZVs}\\
\hline\hline
1&0&$-$&$-$ \\ 
2&0&$-$&$-$ \\ 
3&1&$\tau_3$&$\z_{3}$\\ 
4&0&$-$&$-$\\
5&1&$\tau_5$&$\z_{5}$\\ 
6&0&$-$&$-$\\ 
7&1&$\tau_7$&$\z_{7}$\\ 
8&1&$[\tau_5,\tau_3]$&$\z_{3,5}$\\ 
9&1&$\tau_9$&$\z_{9}$\\
10&1&$[\tau_7,\tau_3]$&$\z_{3,7}$\\
11&2&$\tau_{11},\ [\tau_3,[\tau_5,\tau_3]]$&$\z_{11},\ \z_{3,3,5}$ \\[1mm]
12&2&$[\tau_9,\tau_3],\ [\tau_7,\tau_5]$&$\z_{3,9}\ ,\z_{1,1,4,6}$ \\[1mm]
13&3&$\tau_{13},\ [\tau_3,[\tau_7,\tau_3]],\ [\tau_5,[\tau_5,\tau_3]]$&$\z_{13},\ \z_{3,3,7},\ \z_{3,5,5}$\\[1mm]
14&3&$[\tau_{11},\tau_3],\ [\tau_9,\tau_5],\ [\tau_3,[\tau_3,[\tau_5,\tau_3]]]$
&$\z_{3,11},\ \z_{5,9},\ \z_{3,3,3,5}$\\[1mm]
15&4&$\tau_{15},\ [\tau_3,[\tau_9,\tau_3]],\ [\tau_5,[\tau_7,\tau_3]],\ [\tau_7,[\tau_5,\tau_3]]$&$\z_{15},\ \z_{5,3,7},\ \z_{3,3,9},\ \z_{1,1,3,4,6}$\\[2mm]
16&5&$[\tau_{13},\tau_3],\ [\tau_{11},\tau_5],\ [\tau_9,\tau_7],$&$\z_{3,13},\ \z_{5,11},\ \z_{1,1,6,8},$\\
& & $[\tau_3,[\tau_5,[\tau_5,\tau_3]]],\ [\tau_3,[\tau_3,[\tau_7,\tau_3]]]$&$\z_{3,3,3,7},\ \z_{3,3,5,5}$\\[1mm]
17&7&$\tau_{17},\ [\tau_3,[\tau_3,[\tau_3,[\tau_5,\tau_3]]]],\ [\tau_7,[\tau_7,\tau_3]],\ [\tau_5,[\tau_7,\tau_5]],$&
$\z_{17},\ \zeta_{3,3,3,3,5},\ \zeta_{1,1,3,6,6},\ \zeta_{5,5,7},$\\
 & &$[\tau_3,[\tau_{11},\tau_3]],\ [\tau_9,[\tau_5,\tau_3]],\ [\tau_5,[\tau_9,\tau_3]]$&
$\zeta_{3,3,11},\ \zeta_{5,3,9},\ \zeta_{3,5,9}$ \\[2mm]
18&8&$[\tau_{15},\tau_3],\ [\tau_{13},\tau_5],\ [\tau_{11},\tau_7],$&$\zeta_{3,15},\ \zeta_{5,13},\ \zeta_{1,1,6,10},$\\ 
& & $[\tau_5,[\tau_5,[\tau_5,\tau_3]]],\ [\tau_3,[\tau_3,[\tau_7,\tau_5]]],$&$\zeta_{3,5,5,5},\ \zeta_{5,3,3,7},$\\
 & & $[\tau_5,[\tau_3,[\tau_7,\tau_3]]],\ [\tau_3,[\tau_3,[\tau_9,\tau_3]]],\ [\tau_3,[\tau_5,[\tau_7,\tau_3]]]$&$\zeta_{3,3,3,9},\ 
\zeta_{3,5,3,7},\ \zeta_{1,1,3,3,4,6}$\\[2mm]
19&11&$\tau_{19},\ [\tau_3,[\tau_{13},\tau_3]],\ [\tau_7,[\tau_9,\tau_3]],\ [\tau_9,[\tau_7,\tau_3]],  $&$\z_{19},\ \z_{3,3,13},\ \z_{7,3,9},\ \z_{1,1,3,6,8},$\\
& & $\ [\tau_5,[\tau_{11},\tau_3]],\ [\tau_{11},[\tau_5,\tau_3]],\ [\tau_5,[\tau_9,\tau_5]],\ [\tau_7,[\tau_7,\tau_5]],$&$\z_{5,3,11},\ \z_{3,5,11},\ \z_{5,5,9},\ \z_{1,1,5,6,6},$\\
& & $[\tau_3,[\tau_3,[\tau_5,[\tau_5,\tau_3]]]],\ [\tau_5,[\tau_3,[\tau_3,[\tau_5,\tau_3]]]],$&$\z_{3,3,5,3,5},\ 
\z_{3,3,3,5,5},$\\
& & $[\tau_3,[\tau_3,[\tau_3,[\tau_7,\tau_3]]]]  $&$\z_{3,3,3,3,7}$\\[2mm]
20&13&$[\tau_{17},\tau_3],\ [\tau_{15},\tau_5],\ [\tau_{13},\tau_7],\ [\tau_{11},\tau_9],$&$\zeta_{7,13},\ \zeta_{5,15},\ \zeta_{3,17},\ \zeta_{1,1,8,10},$\\ 
& & $[\tau_3,[\tau_3,[\tau_3,[\tau_3,[\tau_5,\tau_3]]]]],$&$\zeta_{3,3,3,3,3,5},$ \\
& & $[\tau_5,[\tau_5,[\tau_7,\tau_3]]],\ [\tau_3,[\tau_5,[\tau_7,\tau_5]]],\ [\tau_3,[\tau_3,[\tau_9,\tau_5]]],$&$\zeta_{5,5,3,7},\ \zeta_{3,5,5,7},\ \zeta_{5,3,3,9},$\\
& & $[\tau_3,[\tau_7,[\tau_7,\tau_3]]],\ [\tau_3,[\tau_5,[\tau_9,\tau_3]]],\ [\tau_3,[\tau_3,[\tau_{11},\tau_3]]],$&$
\zeta_{3,3,7,7},\ \zeta_{3,5,3,9},\ \zeta_{3,3,3,11},$\\
& & $[\tau_5,[\tau_3,[\tau_7,\tau_5]]],\ [\tau_5,[\tau_3,[\tau_9,\tau_3]]]$&$\zeta_{1,1,3,3,4,8},\ \zeta_{1,1,5,3,4,6},$\\[2mm]
21&17&$\tau_{21},\ [\tau_9,[\tau_9,\tau_3]],\ [\tau_7,[\tau_{11},\tau_3]],\ [\tau_7,[\tau_9,\tau_5]],$&$\z_{21},\ \z_{3,9,9},\ \z_{1,1,3,6,10},\ \z_{7,5,9},$\\ 
& & $\ [\tau_5,[\tau_{13},\tau_3]],\ [\tau_5,[\tau_{11},\tau_5]],\ [\tau_5,[\tau_9,\tau_7]],$&$\z_{5,3,13},\ \z_{1,1,5,4,10},\ \z_{1,1,5,6,8},$ \\
& & $[\tau_3,[\tau_{15},\tau_3]],\ [\tau_3,[\tau_{13},\tau_5]],\ [\tau_3,[\tau_{11},\tau_7]],$&$\z_{3,3,15},\ \z_{3,5,13},\ \z_{7,3,11},$\\
& & $[\tau_7,[\tau_3,[\tau_3,[\tau_5,\tau_3]]]],\ [\tau_5,[\tau_3,[\tau_5,[\tau_5,\tau_3]]]],\ $&$\z_{3,5,3,3,7},\ \z_{3,5,3,5,5},$\\
& & $ [\tau_5,[\tau_3,[\tau_3,[\tau_7,\tau_3]]]],\ [\tau_3,[\tau_5,[\tau_5,[\tau_5,\tau_3]]]],$&$\z_{5,3,3,3,7},\ \z_{3,3,5,5,5},$\\
& & $\ [\tau_3,[\tau_3,[\tau_3,[\tau_7,\tau_5]]]],\  [\tau_3,[\tau_5,[\tau_3,[\tau_7,\tau_3]]]],$&$\z_{3,3,5,3,7},\ \z_{1,1,3,3,3,4,6},$\\
& & $\ [\tau_3,[\tau_3,[\tau_3,[\tau_9,\tau_3]]]]$&$\z_{3,3,3,3,9}$\\
22&21&$[\tau_{19},\tau_3],\ [\tau_{17},\tau_5],\ [\tau_{15},\tau_7],\ [\tau_{13},\tau_9]$ & $\z_{3,19},\ \z_{5,17},\ \z_{7,15},\ 
\z_{1,1,8,12}$\\
& & $[\tau_3, [\tau_3, [\tau_{13}, \tau_3]]] $&$\z_{3,3,3,13}$\\
& & $[\tau_{11}, [\tau_3, [\tau_5, \tau_3]]],\ [\tau_3, [\tau_5, [\tau_{11}, \tau_3]]],\ [\tau_3, [\tau_{11}, [\tau_5, \tau_3]]]$&
$ \z_{5,3,3,11},\ \z_{3,5,3,11},\ \z_{3,3,5,11}$\\
& & $[\tau_9, [\tau_3, [\tau_7, \tau_3]]],\ [\tau_7, [\tau_3, [\tau_9, \tau_3]]] ,\ [\tau_3, [\tau_9, [\tau_7, \tau_3]]]  $&$\z_{3,7,3,9},\ 
\z_{1,1,7,3,2,8},\ \z_{1,1,3,3,6,8}$\\
& & $[\tau_7, [\tau_5, [\tau_7, \tau_3]]] ,\  [\tau_7, [\tau_7, [\tau_5, \tau_3]]]] ,\ [\tau_5, [\tau_7, [\tau_7, \tau_3]]]]$&$  \z_{5,3,7,7},\ 
\z_{3,7,5,7},\ \z_{1,1,3,5,6,6}$\\
& & $[\tau_5, [\tau_9, [\tau_5, \tau_3]]],\  [\tau_5, [\tau_5, [\tau_9, [\tau_3]]]]],\ [\tau_3, [\tau_5, [\tau_9, \tau_5]]]$& $\z_{5,3,5,9},\ 
\z_{3,5,5,9},\ \z_{5,5,3,9}$\\
& & $[\tau_5, [\tau_5, [\tau_7, \tau_5]]]$& $\z_{1,1,5,5,4,6}$\\
& & $[\tau_3,[\tau_3,[\tau_3,[\tau_5,[\tau_5,\tau_3]]]]],\ [\tau_3,[\tau_5,[\tau_3,[\tau_3,[\tau_5,\tau_3]]]]]$ & $\z_{3,3,3,5,3,5},\ \z_{3,3,3,3,5,5}$\\
& & $[\tau_3,[\tau_3,[\tau_3,[\tau_3,[\tau_7,\tau_3]]]]]$ & $\z_{3,3,3,3,3,7}$\\
\hline
\end{tabular}
\end{table}
\noindent
Alternatively, we have \cite{YI1}
\be\label{NICE2}
\dim(\Fc_m)=\fc{1}{m}\ \sum_{d|m} \mu\lf(\fc{m}{d}\ri)\ \lf(\sum_{i=1}^3\al_i^d-1-(-1)^d\ri)\ ,
\ee
with $\al_i$ being the three roots of the cubic equation $\al^3-\al-1=0$.
The graded space of irreducible (primitive) MZVs
$Z=\tfrac{\Zc_{>0}}{\Zc_{>0}\cdot\Zc_{>0}}$ with $\Zc_{>0}=\oplus_{w>0}\Zc_w$ is isomorphic to the dual of 
$\Fc$, i.e.~$\dim(Z_m)=\dim(\Fc_m)$ \cite{Goncharov1,Goncharov2}.
This property relates linearly independent elements $\Fc$ in the $\ap$--expansion of \req{Observation} or \req{Fmotivic} to primitive MZVs.
The linearly independent algebra elements of $\Fc$ and irreducible (primitive) MZVs (in lines of \cite{DataMine})  at each weight $m$ are displayed in the Table~\ref{tabs} through weight $m=22$.

The  generators $M_{2n+1}$ defined in \req{PM} are represented as $(N-3)!\times (N-3)!$--matrices and enter the commutator structure 
\be\label{COMMU}
[M_{n_2},[M_{n_3},\ldots,[M_{n_r},M_{n_1}]]\ldots]
\ee
in the expansion of \req{Observation} or \req{Fmotivic}. These structures can be related to a  graded Lie algebra over~$\IQ$
\be\label{Lalgebra}
\Lc=\bigoplus_{r\geq 1} \Lc_{r}\ ,
\ee
which is generated by the symbols $M_{2n+1}$ with the
Lie bracket $(M_i,M_j)\mapsto[M_i,M_j]$. The grading is defined by assigning $M_{2n+1}$
the degree $2n+1$.  More precisely, the algebra $\Lc$ is generated by the following elements:
\be\label{ALGEBRA}
M_3,\ M_5,\ M_7,\ [M_5,M_3],\ M_9,\ [M_7,M_3],\ M_{11},\ [M_3[M_5,M_3]],\ [M_9,M_3],\ [M_7,M_5],\ldots\ .
\ee
However, this Lie algebra $\Lc$ is not free for generic matrix representations $M_{2i+1}$ referring to any $N\geq 5$. Hence, generically $\Lc \not\simeq \Fc$.
In fact, for $N=5$ at weight $w=18$ we find the relation $[M_3,[M_5,[M_7,M_3]]]=[M_5,[M_3,[M_7,M_3]]]$
leading to $\dim(\Lc_{18})=7$ in contrast to  $\dim(\Fc_{18})=8$.

For a given multiplicity $N$ the generators $M_{2n+1}$, which are represented as $(N-3)!\times (N-3)!$--matrices, 
are related to their transposed $M_l^t$ by a similarity (conjugacy) transformation $S_0$
\be\label{conjugate}
S_0^{-1}\;M_l^t\;S_0=M_l\ ,
\ee
i.e.~$M_l$ and $M_l^t$ are similar (conjugate) to each other. The matrix $S_0$ is symmetric and 
has been introduced in \cite{SS}. 
The relation \req{conjugate} implies, that the matrices $M_l$ are 
conjugate to symmetric matrices. An immediate consequence is 
the set of relations
\be\label{vanishiii}
S_0\ \Qc_{(r)} +(-1)^r\ \Qc^t_{(r)} S_0=0\ ,
\ee
for  any nested commutator of generic depth $r$
$$\Qc_{(r)}=[M_{n_2},[M_{n_3},\ldots,[M_{n_r},M_{n_1}]]\ldots]\;,\ \ \ \ \ 
r\geq2 \ .$$
As a consequence any commutator $\Qc_{2r}$ is
similar to an anti--symmetric matrix and any commutator $\Qc_{2r+1}$ is
similar to a symmetric and traceless matrix.
Depending on the multiplicity~$N$ the relations \req{vanishiii} impose constraints on the number of 
independent generators at a given weight $m$ given  in the Table 1. 
E.g.~for $N=5$ the constraints \req{vanishiii} imply\footnote{
The relation \req{vanishiii} implies, that any commutator $\Qc_{(2)}$ is
similar to an anti--symmetric matrix, and hence \req{generalcomm} implies 
$[\; [M_a,M_b],[M_c,M_d]\; ]=0$, which in turn as a result of the Jacobi relation yields the following identity:
$[M_a,[M_b,[M_c,M_d]]]-[M_b,[M_a,[M_c,M_d]]] = [\; [M_a,M_b],[M_c,M_d]\; ]=0$.
Furthermore, \req{vanishiii} implies that the commutator $\Qc_{(3)}$ is similar to a symmetric and traceless matrix.
As a consequence from \req{generalcomm1}, we obtain the following anti--commutation relation:
$\{\; [M_a,M_b],[M_c,[M_d,M_e]\; \}=0$. Relations for $N=5$ between different matrices $M_{2i+1}$ have also been discussed in \cite{Boels:2013jua}.}:
\begin{eqnarray}
r_1+r_2 \in 2\IZ^+&&:\ \ \ [\; \Qc_{(r_1)},\tilde \Qc_{(r_2)}\; ]=0\ ,\label{generalcomm}\\
r_1+r_2 \in 2\IZ^++1&&:\ \ \{\; \Qc_{(r_1)},\tilde \Qc_{(r_2)}\; \}=0\ .\label{generalcomm1}
\end{eqnarray}
As a consequence, for $N=5$
the number of independent elements at a given weight $m$ does not agree with the formulae 
\req{NICE1} nor \req{NICE2} starting at weight $w=18$. The actual number of independent commutator structures
at weight $w$ is depicted in Table~\ref{tab2}.
\begin{table}[H]
\caption{Linearly independent elements in $\Fc, \Lc^{(5)}_m$ and primitive MZVs for $m=18,\ldots,23$ for $N=5$.}\label{tab2}
\hskip3.25cm
\begin{tabular}{|c|c|c|c|}
\hline
$m$			
&$\dim(\Fc_m)$
&$\dim(\Lc^{(5)}_m)$
&{\rm irreducible MZVs}\\
\hline\hline
18&8&7&7\\ 
19&11&11&11\\ 
20&13&11&11\\ 
21&17&16&16\\
22&21 &16 &16 \\ 
23&28 &25 &25 \\ 
\hline
\end{tabular}
\end{table}
\noindent
Therefore, for $N=5$ an other algebra $\Lc^{(5)}$ rather than $\Fc$ is relevant for describing the expansion of \req{Observation} or \req{Fmotivic}. For $N\geq6$ we expect the mismatch 
$\dim(\Fc_m)\neq \dim(\Lc_m)$ to show up at higher weights $m$.
This way, for each $N\geq 5$ we obtain a different algebra $\Lc^{(N)}$, which is not free.
However, we speculate that for $N$ large enough, the matrices $M_{2k+1}$ should give rise
to the free Lie algebra $\Fc$, i.e.:
\begin{svgraybox}
\be
\lim_{N\ra\infty}\Lc^{(N)}\simeq\Fc\ .
\ee
\end{svgraybox}

\section{Open and closed superstring amplitudes}
\label{Physics}

\begin{figure}[H]
\caption{World--sheet describing the scattering of $N$ open strings.}
\centering
\includegraphics[width=0.95\textwidth]{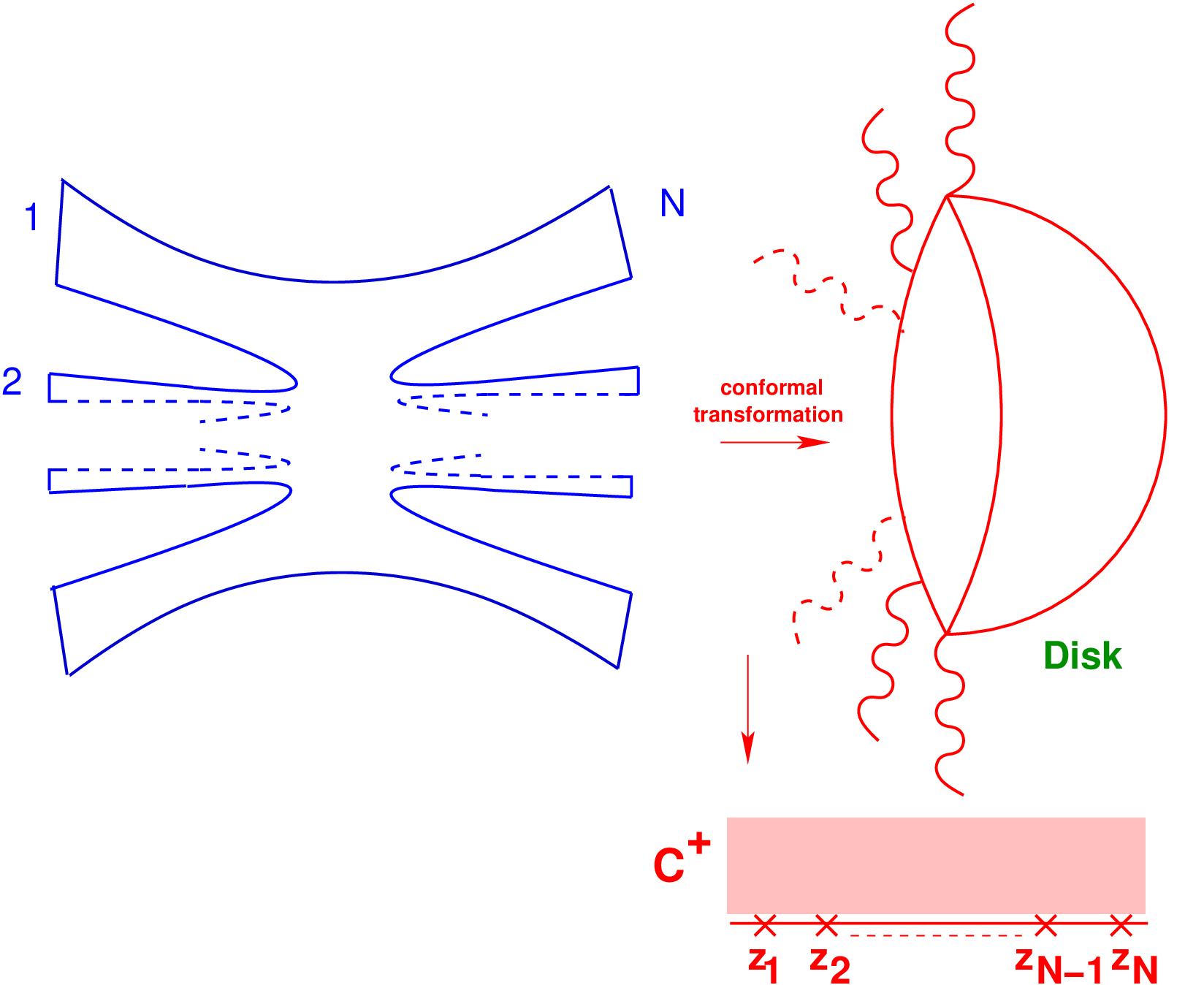}
\label{S6en}
\end{figure}

The world--sheet describing the tree--level scattering of $N$ open strings is depicted in Fig.~\ref{S6en}. 
\noindent
Asymptotic scattering of strings yields the string $S$--matrix defined by the emission and absorption
of strings at space--time infinity, i.e.~the open strings are incoming and outgoing at infinity. In this case the world--sheet can conformally be mapped to the half--sphere 
with the emission and absorption of strings taking place at the boundary through some vertex operators. Source boundaries 
representing the emission and absorption of strings at infinity become points accounting for the vertex operator insertions 
along the boundary of the half--sphere (disk).
After projection onto the 
upper half plane $\IC^+$ the strings are created at the $N$ positions $z_i,\ i=1,\ldots,N$ along the 
(compactified) real axis $\IR\IP^1$. By this there appears a natural ordering $\Pi\in S_N$ of open string vertex operator insertions $z_i$  along the boundary of the disk given by $z_{\Pi(1)}<\ldots<z_{\Pi(N)}$.
To conclude, the topology of the string world--sheet describing tree--level scattering of open strings is
a disk or upper half plane $\IC^+$. 
On the other hand, the tree--level scattering of closed strings is  characterized by a complex sphere
 ${\bf P}^1$ with vertex operator insertions on it.

At the $N$ positions $z_i$ massless strings carrying the external four--momenta $k_i,\ i=1,\ldots,N$ and other quantum numbers are created, subject to momentum conservation $k_1+\ldots+k_N=0$. 
Due to conformal invariance one has to integrate over all vertex operator positions $z_i$ in any amplitude computation. Therefore, for a given ordering $\Pi$ open string amplitudes $\Ac^{o}(\Pi)$ are expressed by integrals along the boundary of the world--sheet disk (real projective line) as
iterated (real) integrals on $\IR\IP^1$ giving rise to multi--dimensional integrals on the space $\Mo(\IR)$  defined in \req{StringDisk}.
The $N$ external four--momenta $k_i$ constitute the kinematic invariants of the scattering process:
\be\label{Mandel}
s_{ij}=(k_i+k_j)^2=2 k_ik_j\ .
\ee
Out of \req{Mandel} there are  $\h N(N-3)$ independent kinematic invariants involving $N$ external momenta $k_i,\ i=1,\ldots,N$.
Any amplitude analytically depends on those independent kinematic invariants $s_{ij}$.

A priori there are $N!$ orderings $\Pi$ of the vertex operator positions $z_i$ along the boundary.
However, string world--sheet symmetries like cyclicity, reflection and  parity give relations between different orderings. In fact, by using monodromy properties on the world--sheet further relations
are found and  any superstring amplitude $\Ac^{o}(\Pi)$
of a given ordering $\Pi$ can be expressed in terms of a minimal basis of $(N-3)!$ amplitudes \cite{Stieberger:2009hq,Vanhove}:
\be\label{defA}
\Ac^{o}(\si):=\Ac^{o}(1,\si(2,\ldots,N-2),N-1,N)\ \ \ ,\ \ \ \si\in S_{N-3}\ .
\ee
The amplitudes \req{defA} are functions of the string tension $\ap$. Power series expansion in $\ap$  yields iterated integrals \req{MZVPeriod} multiplied by some polynomials in the parameters 
\req{Mandel}.

On the other hand, closed string amplitudes  are given by integrals over the complex world--sheet sphere 
$\IP^1$ as iterated integrals  integrated independently on all choices of paths.
While in the $\ap$--expansion of open superstring tree--level amplitudes generically the whole space of MZVs \req{MZV} enters  \cite{OprisaWU,GRAV,SS}, closed superstring tree--level amplitudes exhibit only a subset of MZVs  appearing in their $\ap$--expansion \cite{GRAV,SS}. 
This subclass can be identified
\cite{StiebergerWEA} as the single--valued multiple zeta values (SVMZVs) \req{trivial}.

The open superstring $N$--gluon tree--level amplitude $\Af^{o}_N$ in type I superstring theory
decomposes into a sum
\be\label{DECO}
{\frak A}^{o}_N=(g_{YM}^{o})^{N-2}\ \sum_{\Pi\in S_{N}/{\bf Z}_2}\Tr(T^{a_{\Pi(1)}}\ldots T^{a_{\Pi(N)}})\
\Ac^{o}(\Pi(1),\ldots,\Pi(N))
\ee
over color ordered subamplitudes $\Ac^{o}(\Pi(1),\ldots,\Pi(N))$ supplemented by a group trace
 over matrices $T^a$ in the fundamental representation. Above, the YM coupling is denoted by $g_{YM}^{o}$, which
in type I superstring theory is given by $g_{YM}^{o}\sim e^{\Phi/2}$ with the dilaton field $\Phi$.
The sum runs over all permutations $S_N$ of labels $i=1,\ldots,N$
modulo cyclic permutations ${\bf Z}_2$, which preserve the group trace.
The $\ap\ra0$ limit of the open superstring amplitude \req{DECO}
matches the $N$--gluon scattering amplitude of super Yang--Mills (SYM):
\be\label{FTlimes}
\lf.\Ac^o(\Pi(1),\ldots,\Pi(N))\ri|_{\ap=0}=A(\Pi(1),\ldots,\Pi(N))\ .
\ee
As a consequence from \req{defA} also in SYM one
has a minimal basis of $(N-3)!$ independent partial subamplitudes \cite{Bern:2008qj}:
\be\label{subYM}
A(\si):=A(1,\si(2,\ldots,N-2),N-1,N)\ \ \ ,\ \ \ \si\in S_{N-3}\ .
\ee
Hence, for the open superstring amplitude we may consider a vector $\Ac^{o}$ with its entries
$\Ac^{o}_\si=\Ac^{o}(\si)$  describing the
$(N-3)!$ independent open $N$--point superstring subamplitudes \req{defA}, while for SYM
we have another vector  $A$ with entries $A_{\si}=A(\si)$:
\begin{eqnarray}
\Ac^o&=&(N-3)!\ \mbox{dimensional vector encompassing}\nonumber\\ 
&&\mbox{\hskip1cm {\bf all} independent superstring subamplitudes}
\ \Ac^o_\sigma=\Ac^o(\sigma),\ \sigma\in S_{N-3}\ ,\nonumber\\[3mm]
A&=&(N-3)!\ \mbox{dimensional vector encompassing}\nonumber\\
&&\mbox{\hskip1cm  {\bf all} independent SYM subamplitudes}
\ A_\sigma=A(\sigma),\ \sigma\in S_{N-3}\ .\nonumber
\end{eqnarray}
The two linear independent $(N-3)!$--dimensional vectors $\Ac$ and $A$ are related by a non--singular matrix of rank $(N-3)!$. An educated guess is the following relation
\be\label{Obvious}
\Ac^o=F\ A\ ,
\ee
with the period matrix $F$ given in \req{Period}. 
Note, that with \req{FFT} the Ansatz \req{Obvious} matches the condition \req{FTlimes}.
In components the relation \req{Obvious} reads:
\be\label{OPEN}
\Ac^{o}(\pi)=\sum_{\sigma\in S_{N-3}}
F_{\pi\sigma}\ A(\sigma)\ \ \ ,\ \ \ \pi\in S_{N-3}\ .
\ee
In fact, an explicit string computation proves the relation \req{Obvious} \cite{MafraNV,MafraNW}. 

\begin{figure}[H]
\caption{$N$--gons describing the cyclic structures of holomorphic and anti--holomorphic
cell forms.}
\centering
\includegraphics[width=0.4\textwidth]{polygon2.eps}\includegraphics[width=0.4\textwidth]{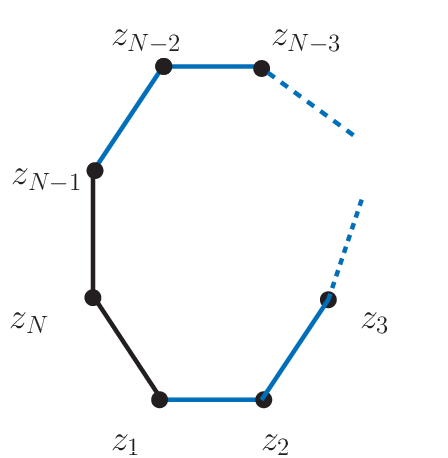}
\label{holoanti}
\end{figure}

Let us now move on to the scattering of closed strings. In heterotic string vacua gluons are described by massless
closed strings. Therefore, we shall consider the closed superstring $N$--gluon tree--level 
amplitude $\Af_{N}^{c}$ in heterotic superstring theory.
The string world--sheet describing the tree--level scattering of $N$
closed strings has the topology of a complex sphere with $N$ insertions
of  vertex operators. The closed string has holomorphic and anti--holomorphic fields. The anti--holomorphic part is similar to the open string case and  describes the space--time (or superstring) part. On the other hand, the holomorphic part accounts for the gauge degrees of freedom through current insertions on the world--sheet. 
As in the open string case \req{DECO}, the single trace part decomposes into the sum
\be\label{TODOst}
\Af_{N,\;{\rm s.t.}}^{c}=(g_{YM}^{\rm HET})^{N-2}\ \sum_{\Pi\in S_{N}/{\bf Z}_2}{\rm tr}(T^{a_{\Pi(1)}}\ldots T^{a_{\Pi(N)}})\ \Ac^{c}(\Pi(1),\ldots,\Pi(N))
\ee
over partial subamplitudes $\Ac^{c}(\Pi)$ times a group trace  over matrices $T^a$ in the vector representation. 
In the $\ap\ra0$ limit the latter match the $N$--gluon scattering subamplitudes of SYM
\be\label{FTlimesC}
\lf.\Ac^c(\Pi(1),\ldots,\Pi(N))\ri|_{\ap=0}=A(\Pi(1),\ldots,\Pi(N))\ ,
\ee
similarly to open string case \req{FTlimes}.
Again, the partial subamplitudes $\Ac^{c}(\Pi)$ can be expressed in terms of a minimal basis of $(N-3)!$ elements $\Ac^{c}(\rho),\ \rho\in S_{N-3}$. The latter have been computed in \cite{Stieberger:2014hba}
and are given by
\be\label{intresult}
\Ac^{c}(\rho)=(-1)^{N-3}\ \sum_{\sigma\in S_{N-3}}\sum_{\ov\rho\in S_{N-3}}
J[\rho\,|\,\ov\rho]\ S[\ov\rho|\si]\ A(\si)\ ,\ \ \rho\in S_{N-3}\ ,
\ee
with the complex sphere integral\footnote{The factor $V_{\rm CKG}$ accounts for the volume of
the conformal Killing group of the sphere after choosing the conformal gauge. It will be
canceled by fixing three vertex positions according to \req{FIX} and introducing the respective $c$--ghost factor $|z_{1,N-1}z_{1,N}z_{N-1,N}|^2$.} 
\begin{eqnarray}
 J[\rho\,|\,\ov\rho]&:=&V_{\rm CKG}^{-1}\  \lf(\prod_{j=1}^{N}\int\limits_{z_j\in\IC} d^2z_j\ri)
\ \prod\limits_{i<j}^{N} |z_{ij}|^{2\ap s_{ij}}\ \fc{1}{z_{1\rho(2)}z_{\rho(2),\rho(3)}\ldots z_{\rho(N-2),N-1}z_{N-1,N}z_{N,1}}\cr
&\times&\displaystyle{\fc{1}{ \ov z_{1\ov\rho(2)} \ov z_{\ov\rho(2),\ov\rho(3)} \ldots \ov z_{\ov\rho(N-3),\ov\rho(N-2)}\ov z_{\ov\rho(N-2), N}\ov z_{N,N-1}\ \ov z_{N-1,1} }}\ ,\label{sphereH}
\end{eqnarray}
the kernel $S$ introduced  in \req{kernel} and the SYM amplitudes \req{subYM}.
In \req{sphereH} the rational function comprising the dependence on holomorphic and anti--holomorphic vertex operator  positions shows some pattern depicted in Fig.~\ref{holoanti}.
\noindent
Based on the results \cite{StiebergerWEA} the following (matrix) identity has been established in \cite{Stieberger:2014hba}
\begin{svgraybox}
\be\label{NICE}
J=\sv(Z)\ ,
\ee
\end{svgraybox}
\noindent
relating the complex integral \req{sphereH} to the real iterated integral \req{ZZ}.
The holomorphic part of \req{sphereH} simply turns into the corresponding integral ordering of
\req{ZZ}.
As a consequence of \req{NICE} we find  the following relation between the closed \req{intresult} and open  \req{OPEN} superstring gluon amplitude \cite{Stieberger:2014hba}:
\be\label{STRIKREL}
\Ac^{c}(\rho)=\sv \lf(\Ac^{o}(\rho)\ri)\ \ \ ,\ \ \ \rho\in S_{N-3}\ .
\ee
To conclude, the single trace heterotic gauge amplitudes $\Ac^{c}(\rho)$ referring to the color ordering 
$\rho$ are simply obtained from the relevant open
string gauge amplitudes $\Ac^{o}(\rho)$ by imposing the projection $\sv$ introduced in \req{mapSV}.
Therefore, the $\ap$--expansion of the heterotic amplitude $\Ac^{c}(\rho)$ can be obtained from that of the open superstring amplitude $\Ac^{o}(\rho)$
by simply replacing MZVs by their corresponding SVMZVs according to the rules 
introduced in \req{mapSV}.
The relation \req{STRIKREL} between the heterotic gauge amplitude $\Ac^{c}$ and the type I gauge amplitude $A^{o}$ establishes a non--trivial relation between closed string and open string 
amplitudes: the $\ap$--expansion  of the closed superstring amplitude can be cast into the same algebraic form as the open superstring amplitude: the closed superstring amplitude  is essentially the single--valued (\sv) version of the open superstring amplitude.

Also closed string  amplitudes other than the heterotic (single--trace) gauge  amplitudes \req{intresult}  can be expressed as single--valued image of some open string amplitudes.
From \req{NICE} the closed string analog of \req{Z0} follows:
\be\label{Wehave}
\lf.J\ri|_{\ap=0}=(-1)^{N-3}\ S^{-1}\ .
\ee
Hence, the set of complex world--sheet sphere integrals \req{sphereH} are the closed string analogs of the open string world--sheet disk integrals \req{ZZ} and serve as building blocks to construct any closed string amplitude. 
After applying partial integrations to remove double poles, which are responsible for 
spurious tachyonic poles, further performing partial fraction decompositions and partial integration relations all closed superstring
amplitudes can be expressed in terms of the basis \req{sphereH}, which in turn through \req{NICE} can be
related to the basis of open string amplitudes \req{ZZ}. As a consequence the $\ap$--dependence
of any closed string amplitude is given by that of the underlying open string amplitudes.
This  non--trivial connection between open and closed string  amplitudes  at the string tree--level  points into a deeper connection between gauge and gravity amplitudes than what is implied by Kawai--Lewellen--Tye relations \cite{KawaiXQ}.

\section{Complex  vs. iterated integrals}

Perturbative open and closed string amplitudes seem to be rather different due to their underlying different world--sheet topologies with or without boundaries, respectively. On the other hand,
mathematical methods entering their computations reveal some unexpected connections.
As we have seen in the previous section a new relation \req{NICE} between open \req{ZZ} and closed \req{sphereH} string world--sheet integrals holds.

Open string world--sheet disk integrals \req{ZZ} are described as real iterated integrals on the space 
$\Mo(\IR)$ defined in \req{StringDisk}, while closed string world--sheet sphere integrals \req{sphereH} are
given by integrals on the space $\Mo(\IC)$ defined in \req{MSP}. The latter integrals, which can be considered as iterated integrals on $\IP^1$ integrated independently on all choices of paths,
are more involved than the real iterated integrals appearing in open string amplitudes.
The observation \req{NICE} that complex integrals can be expressed as real iterated integrals subject to the projection $\sv$ has exhibited non--trivial relations between open and closed string amplitudes \req{STRIKREL}.
In this section we shall elaborate on these connections at the level of the world--sheet integrals.

\begin{figure}[H]
\centering
\includegraphics[width=0.35\textwidth, height=100px]{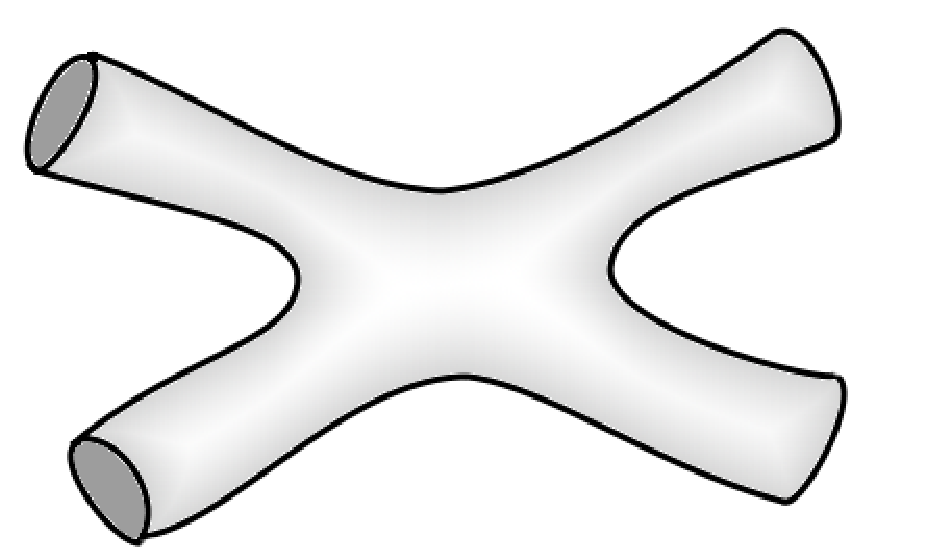}
\hskip1.5cm
\includegraphics[width=0.35\textwidth]{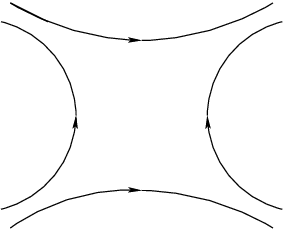}
\caption{Relation between closed and open string world--sheet diagram describing four--point scattering.}
\label{SVPIC}
\end{figure}

The simplest example of \req{NICE} arises for $N=4$ yielding the relation
\be\label{simple}
\int_\IC d^2z\ \fc{|z|^{2 s}\ |1-z|^{2 u}}{z\ (1-z)\ \ov z}=
\sv\lf(\int_0^1 dx\ x^{ s-1}\ (1-x)^{ u}\ri)\ ,
\ee
with $s, u\in \IR$ such that both integrals converge.
While the integral on the l.h.s.~of \req{simple} describes a four--point closed string amplitude
the integral on the r.h.s.~describes a four--point open string amplitude. Hence,  
the meaning of \req{simple} w.r.t.~to the corresponding closed vs. open string world--sheet 
diagram describing  four--point scattering \req{STRIKREL} can be depicted as  Fig.~\ref{SVPIC}.

\vskip-3.75cm

\hskip5cm {\Large =\ sv}
\vskip3.25cm

\noindent
After performing the integrations the relation \req{simple} becomes (with $s+t+u=0$):
\be
\fc{\Gamma(s)\ \Gamma(u)\ \Gamma(t)}{\Gamma(-s)\ \Gamma(-u)\ \Gamma(-t)}=\sv\lf(\fc{\Gamma(1+s)\ \Gamma(1+u)}{\Gamma(1+s+u)}\ri)\ .
\ee
Essentially, this equality (when acting on $[e_0,e_1]$) represents the relation between the Deligne \req{DeligneA} and Drinfeld \req{Drinfeld} associators in the explicit representation of the limit 
${\rm mod} (g')^2$ with $(g')^2=[g,g]^2$,\ $s=-{\rm ad}_{e_1},\ u={\ad}_{e_0}$ and ${\rm ad}_x y=[x,y]$, i.e.~dropping all quadratic commutator terms
\cite{Drummond:2013vz,StiebergerWEA}.
Note, that applying Kawai--Lewellen--Tye (KLT) relations \cite{KawaiXQ} to the complex integral of \req{simple} rather yields
\be\label{KLTEx}
\int_\IC d^2z\ \fc{|z|^{2s}\ |1-z|^{2u}}{z\ (1-z)\ \ov z}=\sin(\pi u)\ 
\lf(\int_0^1 dx\ x^{s-1}\ (1-x)^{u-1}\ri)\ 
\lf(\int_1^\infty dx\ x^{t-1}\ (1-x)^{u}\ri)\ ,
\ee
expressing the  latter in terms of a square of real iterated integrals instead of  a 
single real iterated integral as in \req{simple}. In fact, any direct computation of this complex integral
 by means of a Mellin representation or Gegenbauer decomposition  ends up at \req{KLTEx}.

Similar \req{simple} explicit and direct correspondences \req{NICE} between the
complex sphere integrals $J$ and the real disk integrals $Z$ can be made for $N\geq 5$.
In order to familiarize with the matrix notation let us explicitly write the case \req{NICE}
for $N=5$ (with \req{FIX}):
\begin{eqnarray}
&\hskip-1.25cm
\lf(\begin{array}{lcl}
\displaystyle{\int\limits_{z_2,z_3\in\IC}d^2z_2\ d^2z_3\ \fc{\prod\limits_{i<j}^4|z_{ij}|^{2\ap s_{ij}}}{z_{12}z_{23}z_{34}\ \ov z_{12}\ov z_{23}}}&
\displaystyle{\int\limits_{z_2,z_3\in\IC}d^2z_2\ d^2z_3\ \fc{\prod\limits_{i<j}^4|z_{ij}|^{2\ap s_{ij}}}{z_{12}z_{23}z_{34}\ \ov z_{13}\ov z_{32}}}\\
\displaystyle{\int\limits_{z_2,z_3\in\IC}d^2z_2\ d^2z_3\ \fc{\prod\limits_{i<j}^4|z_{ij}|^{2\ap s_{ij}}}{z_{13}z_{32}z_{24}\ \ov z_{12}\ov z_{23}}}&
\displaystyle{\int\limits_{z_2,z_3\in\IC}d^2z_2\ d^2z_3\ \fc{\prod\limits_{i<j}^4|z_{ij}|^{2\ap s_{ij}}}{z_{13}z_{32}z_{24}\ \ov z_{13}\ov z_{32}}}
\end{array}\ri)\nonumber\\[5mm]
&\hskip0.25cm=
\sv\lf(\begin{array}{lcl}
\displaystyle{\int\limits_{0<z_2<z_3<1}dz_2\ dz_3\ \fc{\prod\limits_{i<j}^4|z_{ij}|^{\ap s_{ij}}}{z_{12}z_{23}}}&
\displaystyle{\int\limits_{0<z_2<z_3<1}dz_2\ dz_3\ \fc{\prod\limits_{i<j}^4|z_{ij}|^{\ap s_{ij}}}{z_{13}z_{32}}}\\
\displaystyle{\int\limits_{0<z_3<z_2<1}dz_2\ dz_3\ \fc{\prod\limits_{i<j}^4|z_{ij}|^{\ap s_{ij}}}{z_{12}z_{23}}}&
\displaystyle{\int\limits_{0<z_3<z_2<1}dz_2\ dz_3\ \fc{\prod\limits_{i<j}^4|z_{ij}|^{\ap s_{ij}}}{z_{13}z_{32}}}
\end{array}\ri)
\label{NICEfive}
\end{eqnarray}
In \req{NICEfive} we explicitly see how the presence of the holomorphic gauge insertion in the complex integrals results in the projection onto real integrals involving only the right--moving part.
Similar matrix relations can be extracted from \req{NICE} beyond $N>5$. 

From \req{NICE} it follows that the $\ap$--expansion of the closed string amplitude can be obtained from that of the open superstring amplitude by simply replacing MZVs by their corresponding SVMZVs according to the rules introduced in \req{mapSV}. Hence, closed string amplitudes use  only  the smaller subspace 
of SVMZVs.
From a physical point of view SVMZVs
appear in the computation of graphical functions (positive functions on the punctured complex
plane) for certain Feynman amplitudes \cite{Schnetz}.
In supersymmetric Yang--Mills  theory a large class of Feynman integrals in four space--time dimensions lives in the subspace of SVMZVs or SVMPs.
As pointed out by Brown in \cite{SVMZV}, this fact opens the interesting possibility
to replace general amplitudes  with their single--valued versions (defined in \req{mapSV} by the map $\sv$),
which should lead to considerable simplifications.
In string theory this simplification occurs by replacing open superstring amplitudes by their
single--valued versions describing closed superstring amplitudes.

\vskip0.5cm
\goodbreak
\leftline{\noindent{\bf Acknowledgments}}

\noindent

We wish to thank the organizers (especially Jos\'e Burgos, 
Kurush Ebrahimi-Fard, and Herbert Gangl) of the workshop {\em Research Trimester on Multiple Zeta Values, Multiple Polylogarithms, and Quantum Field Theory} and the conference {\em  Multiple Zeta Values, Modular Forms
and Elliptic Motives II} at ICMAT, Madrid  for inviting me to present the work exhibited in this publication and creating a stimulating atmosphere. 


%
%
%

\end{document}